\documentclass{article}
\usepackage{amsfonts}
\usepackage{pdflscape}
\usepackage{lscape}
\usepackage{censor}
\usepackage{graphicx}
\usepackage[legalpaper, margin=1in]{geometry}
\usepackage[skip=1pt,font=scriptsize]{caption}
\usepackage[utf8]{inputenc}
\usepackage{hyperref}
\usepackage{authblk}
\usepackage{lineno}
\usepackage[dvipsnames]{xcolor}
\usepackage[normalem]{ulem}
\hypersetup{
    colorlinks=true,
    linkcolor=blue,
    filecolor=red,      
    urlcolor=blue,
}
\usepackage{url}

\title{Functional Connectome Fingerprint Gradients in Young Adults}
\author{Uttara Tipnis$^{1,2,3}$, Kausar Abbas$^{1,2}$, Elizabeth Tran$^{3}$, Enrico Amico$^{4,5}$, Li Shen$^{6}$, Alan D. Kaplan$^{3}$*, Joaquín Goñi$^{1,2,7}$*}
\date{%
$^1$ Purdue Institute for Integrative Neuroscience, Purdue University, West Lafayette, IN, USA \\
$^2$ School of Industrial Engineering, Purdue University, West Lafayette, IN, USA \\
$^3$ Lawrence Livermore National Laboratory, Livermore, California, USA\\
$^4$ Institute of Bioengineering, Center for Neuroprosthetics, École Polytechnique Fédérale de Lausanne, Geneva, Switzerland\\
$^5$ Department of Radiology and Medical Informatics,Université de Genève, Geneva, Switzerland\\
$^6$ Perelman School of Medicine, University of Pennsylvania, PA, USA \\
$^7$ Weldon School of Biomedical Engineering, Purdue University, West Lafayette, IN, USA\\
$^*$ Both authors contributed equally\\
AK: kaplan7@llnl.gov\\
JG: jgonicor@purdue.edu
}

\begin{document}

\maketitle

\section*{Abstract}

  The assessment of brain \textit{fingerprints} has emerged in the recent years as an important tool to study individual differences and to infer quality of neuroimaging datasets. Studies so far have mainly focused on connectivity fingerprints between different brain scans of the same individual. Here, we extend the concept of brain connectivity fingerprints beyond test/retest and assess \textit{fingerprint gradients} in young adults by developing an extension of the differential identifiability framework. To do so, we look at the similarity between not only the multiple scans of an individual (\textit{subject fingerprint}), but also between the scans of monozygotic and dizygotic twins (\textit{twin fingerprint}). We have carried out this analysis on the 8 fMRI conditions present in the Human Connectome Project -- Young Adult dataset, which we processed into functional connectomes (FCs) and timeseries parcellated according to the Schaefer Atlas scheme, which has multiple levels of resolution. Our differential identifiability results show that the fingerprint gradients based on genetic and environmental similarities are indeed present when comparing FCs for all parcellations and fMRI conditions. Importantly, only when assessing optimally reconstructed FCs, we fully uncover fingerprints present in higher resolution atlases. We also study the effect of scanning length on subject fingerprint of resting-state FCs to analyze the effect of scanning length and parcellation. In the pursuit of open science, we have also made available the processed and parcellated FCs and timeseries for all conditions for $\sim$1200 subjects part of the HCP-YA dataset to the scientific community.

\section{Introduction} \label{sec_intro}

    With the advent of improved neuroimaging acquisition techniques, there has been a surge in the availability of high quality neuroimaging data in recent years. Data repositories such as the different Human Connectome Project (HCP) \cite{van2013wu} datasets (HCP Young Adult \cite{van2013wu}, HCP Aging \cite{bookheimer2019lifespan}, HCP Development \cite{somerville2018lifespan}, etc.), 1000 Functional Connectomes Project (\url{http://fcon_1000.projects.nitrc.org/}), UK Biobank \cite{allen2014uk}\cite{miller2016multimodal}, and the Alzheimer's Disease Neuroimaging Initiative (ADNI) \cite{petersen2010alzheimer} among others are openly available to the scientific community. These data repositories, although highly valuable, do not always provide the users with ready-to-use processed subject-level whole-brain functional connectomes (FCs). Instead, they provide raw data or data that has been only minimally processed \cite{glasser2013minimal}\cite{makropoulos2018developing}. Hence, it is typically up to the researcher to estimate single session whole-brain functional connectomes from fMRI and T1 data. This step is critical \cite{power2020critical}\cite{parkes2018evaluation}\cite{power2018ridding}\cite{power2014methods}\cite{power2015recent}\cite{burgess2016evaluation} for subsequent brain connectivity and network neuroscience analyses \cite{fornito2016fundamentals}\cite{sporns2010networks}\cite{sporns2012discovering}. This can be a difficult task due to the knowledge required as well as the amount of computational power necessary to process large datasets such as HCP. \par
    
    These open-source datasets are usually shared with the community with either no or minimal artefact and/or noise removal. This is an efficient and suitable strategy for neuroimaging data sharing. One of the main reasons is that MRI data processing is constantly evolving, with registration, processing and denoising methods constantly evolving \cite{power2020critical}\cite{parkes2018evaluation}\cite{power2018ridding}\cite{power2014methods}\cite{power2015recent}\cite{burgess2016evaluation} as well as new brain atlases \cite{schaefer2018local}\cite{glasser2016multi}\cite{salehi2020there} being provided to the community. Providing raw or minimally processed datasets allows for up-to-date processing techniques to be applied to the dataset. \par

    Amongst the many different choices one has to make while processing raw neuroimaging data to obtain subject-level, single-session, whole-brain functional connectomes, the choice of the parcellation is very important \cite{schaefer2018local}\cite{glasser2016multi}\cite{salehi2020there}. The subsequent analysis of the connectomes depends on the level of granularity of a parcellation. The importance of the parcellation granularity has been shown, for instance, when evaluating brain fingerprints \cite{finn2015functional}\cite{abbas2020regularization}. Schaefer et al., 2018 \cite{schaefer2018local} recently published a scheme of parcellations that gives the user the ability to assess up to 10 different levels of granularity (atlases include 100 to 1,000 brain regions, in steps of 100). Another added advantage of this parcellation scheme is that all ten levels of granularity are further divisions of the resting state functional networks proposed by Yeo et al., 2011 \cite{yeo2011organization}. \par

     There is no standard procedure to decide on the brain parcellation to estimate functional connectomes. This is also true for the artefact and noise removal steps in the fMRI data processing. To that end, the amount of subject fingerprint present in the resultant functional connectomes is a useful proxy, as a whole, of the measure of quality of the experimental design, acquisition parameters, and the ultimate estimation of the functional connectomes. Many recent studies have established that functional connectomes have an individual fingerprint that can be used to identify an individual from a population (a process known as \textit{fingerprinting} or \textit{subject-identification}) \cite{amico2018quest}\cite{abbas2020geff}\cite{finn2015functional}\cite{kaufmann2017delayed}\cite{miranda2014connectotyping}\cite{noble2017influences}\cite{rajapandian2020uncovering}\cite{mars2018connectivity}\cite{abbas2020geff}\cite{hu2020disentangled}\cite{ngo2020connectomic}. Subject-level fingerprints in the FCs have been found to be reliable and reproducible in high quality datasets  \cite{finn2015functional} (e.g., HCP). Moreover, this fingerprint can be improved by using the differential identifiability framework ($\mathbb{I}\mathit{f}$), which relies on performing group-level decomposition into principal components followed by an iterative reconstruction adding components in descending order of explained variance until the differential identifiability score reaches an optimal value \cite{amico2018quest}. Without a high subject-level fingerprint, brain connectomic analyses that are aimed at finding associations between functional connectivity and cognition, behavior, or disease progression are severely compromised \cite{svaldi2019optimizing}\cite{sripada2020boost}.\par

    Fingerprints are not unique to test/retest sessions of the same individuals. Subjects sharing genetics and/or environment are expected to have, to some extent, a fingerprint. In particular, similar to subject-level fingerprint in brain functional connectomes, it has been established that a fingerprint also exists in the FCs of monozygotic (MZ) and dizygotic (DZ) twin subjects \cite{ge2017heritability}\cite{de2008electroencephalographic}\cite{kumar2018multi}\cite{gritsenko2020twin}\cite{colclough2017heritability}\cite{demeter2020functional}, albeit to a lower extent than the subject-level fingerprint. Kumar et al. 2018 \cite{kumar2018multi} have presented a framework based on manifold approximation for generating brain fingerprints from multimodal data using T1/T2-weighted MRI, diffusion MRI, and resting-state fMRI. Their results show a link between amount of fingerprint and genetic proximity as the MZ twins have more prominent fingerprints than DZ or non-twin siblings. Ge et al. 2017 \cite{ge2017heritability} have used a linear mixed effects model to dissociate intra- and inter-subject variation of a phenotype and computed heritability with respect to stable inter-subject variation in fMRI data as the phenotype. Colclough et al. 2018 \cite{colclough2017heritability} have investigated the influence of genetics and common environment on functional connectomes of individuals obtained from fMRI and MEG data in HCP. Demeter et al. 2020 \cite{demeter2020functional} have applied support vector machine classifiers on resting state fMRI to predict retest and co-twin pairs from two twin datasets (adult and pediatric) that include repeat scans. Gritsenko et al. 2020 \cite{gritsenko2020twin} propose a pair-wise twin classification method to identify the zygosity of twin pairs using the resting state fMRI. The latest release of the HCP-YA \cite{glasser2016multi} dataset includes unrelated subjects, as well as subjects that are related to each other, including MZ and DZ twins. This affords us the opportunity to assess brain connectivity fingerprints not only by comparing test and retest functional connectomes of the same subject (\textit{subject-level fingerprint}), but also by comparing the functional connectomes of MZ and DZ twins (\textit{twin fingerprint}) across different fMRI conditions.\par

    The aim of this study is to provide state-of-the-art processed whole-brain, single-session FCs to the scientific community for conducting research in brain connectomics \cite{sporns2012discovering}\cite{fornito2016fundamentals}\cite{rubinov2010complex}. We provide FCs corresponding to all 10 levels of granularity (100 to 1,000) of the Schaefer parcellations. In terms of artefact/noise removal processing steps, we provide FCs at different level of processing/denoising (e.g. with and without global signal regression \cite{murphy2017towards}\cite{liu2017global}\cite{hayasaka2013functional}\cite{xu2018impact}\cite{gotts2013perils}\cite{saad2012trouble}). In addition, we assess the amount of subject-level fingerprint and twin-fingerprint (MZ and DZ) in each fMRI condition (resting-state and 7 tasks), at different levels of granularity of the Schaefer parcellations. We estimate and uncover these fingerprints using an extended version of the differential identifiability framework ($\mathbb{I}\mathit{f}$) \cite{amico2018quest}\cite{bari2019uncovering}. \par

\section{Methods} \label{sec_methods}

    \subsection{The HCP-YA dataset} \label{sec_HCP_data}
    
          The functional MRI data processed as a part of this study is available in the \href{http://www.humanconnectome.org/study/hcp-young-adult}{Human Connectome Project-Young Adult (HCP-YA) repository} \cite{van2013wu}. The HCP-YA data consists of behavioural and 3T MRI data from 1206 healthy young adult subjects collected between August 2012 and October 2015. 3T MR structural scans are available for 1113 subjects, out of which 889 subjects have fully complete data for all four 3T MRI modalities: structural (T1w and T2w) data, resting state fMRI, task fMRI, and high angular resolution diffusion MRI data. The HCP-YA dataset also has extensive family structures, including siblings and twin pairs (monozygotic and dizygotic). All the subjects are within the age range of 22-37 years at the time of scanning. \textit{Table \ref{table_numconns}} summarizes the number of unrelated subjects and monozygotic and dizygotic twin pairs for resting state and all 7 tasks included in HCP-YA. The term \textit{fMRI Condition} would be used to indicate both resting state and tasks that are included in the dataset.
        
        \begin{table}[h]
            \centering
            \begin{tabular}{|l|c|c|c|}
            \hline
            \textbf{fMRI Condition} & \textbf{Unrelated subjects} & \textbf{MZ twin pairs} & \textbf{DZ twin pairs} \\
            \hline
            REST1 & 435 & 131 & 76 \\
            REST2 & 435 & 131 & 76 \\
            EMOTION & 416 & 124 & 70 \\
            GAMBLING & 438 & 135 & 74 \\
            LANGUAGE & 417 & 129 & 72 \\
            MOTOR & 438 & 134 & 76 \\
            RELATIONAL & 414 & 125 & 69 \\
            SOCIAL & 417 & 128 & 71 \\
            WORKING MEMORY & 436 & 133 & 77 \\
            \hline
            \end{tabular}
            \caption{Summary of the number of unrelated subjects, MZ and DZ twin pairs corresponding to each of the fMRI conditions in the HCP-YA dataset}
            \label{table_numconns}
        \end{table}
    
    \subsubsection{HCP-YA fMRI conditions}
    
        We have used the fMRI data from the HCP-YA 1200 subjects release \cite{van2012human}\cite{van2013wu}. The fMRI resting-state data (HCP-YA filenames: rfMRI\_REST1 and rfMRI\_REST2) were acquired in separate sessions on two different days, with two different phase acquisitions (left to right or LR and right to left or RL) per day \cite{van2012human}\cite{van2013wu}\cite{glasser2013minimal}. The seven fMRI tasks are the following: gambling (tfMRI\_GAMBLING), relational (tfMRI\_RELATIONAL), social (tfMRI\_SOCIAL), working memory (tfMRI\_working memory), motor (tfMRI\_MOTOR), language (tfMRI\_LANGUAGE, including both a story-listening and arithmetic task), and emotion (tfMRI\_EMOTION). Two runs (LR and RL) were acquired for each task. Working memory, gambling, and motor task were acquired on the first day, and the other tasks on the second day \cite{glasser2013minimal}\cite{barch2013function}. \textit{Table \ref{table_task_summary}} summarizes the run time and number of frames per condition:
        
        \begin{table}[h!]
            \centering
            \begin{tabular}{|l|c|c|c|}
                \hline
                 \textbf{fMRI Condition} & \textbf{\#Runs} & \textbf{Run time (min:sec)} & \textbf{\#Frames} \\
                 \hline
                 REST1 & 2 & 14:33 & 1,200\\
                 REST2 & 2 & 14:33 & 1,200\\
                 EMOTION & 2 & 2:16 & 176\\
                 GAMBLING & 2 & 3:12 & 253\\
                 MOTOR & 2 & 3:34 & 284\\
                 LANGUAGE & 2 & 3:57 & 316\\
                 RELATIONAL & 2 & 2:56 & 232\\
                 SOCIAL & 2 & 3:27 & 274\\
                 WORKING MEMORY & 2 & 5:01 & 405\\
                 \hline
            \end{tabular}
                \caption{Summary of the number of runs, run time (in minutes and seconds), and number of frames per run for resting state and 7 tasks included in the HCP-YA dataset}
            \label{table_task_summary}
        \end{table}
        
        The following is a brief description of each fMRI condition. More extensive information may be found in the \href{https://www.humanconnectome.org/storage/app/media/documentation/s1200/HCP_S1200_Release_Reference_Manual.pdf}{HCP S1200 Release Reference Manual}.

        \begin{itemize}
            
            \item \textbf{REST:} Resting state fMRI (rs-fMRI) data was acquired in four runs of approximately 15 minutes each, two runs in each session. The subjects were instructed to keep their eyes open with relaxed fixation on a projected bright cross-hair on a dark background presented in a darkened room. Within each session, oblique axial acquisitions alternated between phase encoding in a right-to-left (RL) direction in one run and phase encoding in a left-to-right (LR) direction in the other run. 1200 frames were obtained per run at 720 ms TR.
            
            \item \textbf{EMOTION:} This task was adapted from the one developed by Hariri et al. \cite{hariri2006preference}. Subjects are shown blocks of trials that either ask them to decide which of two faces on the bottom of the screen match the face at the top, or which of two shapes at the bottom match the shape at the top of the screen. The faces have either an angry or fearful expression. Trials are presented in blocks of 6 trials of the same task (face or shape), with the stimulus presented for 2000 ms and a 1000 ms ITI. Each block is preceded by a 3000 ms task cue (“shape” or “face”). Each of the two runs includes three face blocks and three shape blocks, with 8 seconds of fixation at the end of each run. In total, the emotion processing task fMRI had a run duration of 2:16 minutes per run, with 176 frames per run.
            
            \item \textbf{GAMBLING:} This task has been adapted from the one developed by Delgado and Fiez \cite{delgado2000tracking}. Subjects are asked to play a card guessing game wherein they are asked to guess the number on a mystery card in order to win or lose money. Subjects are told that potential card numbers are between 1 and 9 and to indicate whether they think the mystery card number is more or less than 5 by a button press. Feedback to the subject is the actual number on the card and either a green arrow up with "\$1" for reward or red arrow down with "-\$0.5" for loss. If the mystery number is equal to 5, the trial is considered neutral and a grey double headed arrow is shown. The subjects have 1500 ms to respond with button press, followed by 1000 ms of feedback. If the subject responds before the 1500 ms is over, a fixation cross is displayed. The task is presented in blocks of 8 trials that are either mostly reward or mostly loss. In each of the two runs, there are 2 mostly reward and 2 mostly loss blocks, interleaved with 4 fixation blocks (15 seconds each). In total, the gambling task fMRI had a run duration of 3:12 minutes per run, with 253 frames per run.
            
            \item \textbf{LANGUAGE:} This task was developed by Binder el al. \cite{binder2011mapping} and uses the E-prime scripts provided by them. The task consists of two runs that each interleave four blocks of a story task and four blocks of a math task. The lengths of the blocks vary (average of approximately 30 seconds), but the task was designed so that the math task blocks match the length of the story task blocks, with some additional math trials at the end of the task to complete the 3.8 minute run as needed. The story blocks present participants with brief auditory stories (5-9 sentences) adapted from Aesop’s fables, followed by a 2-alternative forced choice question that asks participants about the topic of the story. The math task also presents trials aurally and requires subjects to complete addition and subtraction problems. Participants push a button to select either the first or the second answer from the options presented. The math task is adaptive to try to maintain a similar level of difficulty across participants. In total, the language task fMRI had a run duration of 3:57 minutes per run, with 316 frames per run.
            
            \item \textbf{MOTOR:} This task was adapted from the one developed by Yeo et al., 2011 \cite{yeo2011organization}. In this task, subjects are shown visual cues asking them to either tap their left or right fingers, or squeeze their left or right toes, or move their tongue to map different motor areas in the brain. There are total 10 movements and each movement type lasted 12 seconds, preceded by a 3 second cue. In each of the two runs, there are 13 blocks, with two of tongue movements, four of hand movements (2 right and 2 left), and four of foot movements (2 right and 2 left). In addition, there are three 15-second fixation blocks per run. In total, the motor task fMRI had a run duration of 3:34 minutes per run, with 284 frames per run.
            
            \item \textbf{RELATIONAL:} This task has been adapted from the work done by Smith et al. \cite{smith2007localizing}. The stimuli are six different shapes filled with one out of six different textures. Subjects are presented with 2 pairs of objects, one at the top of the screen and the other at the bottom. They have to first decide whether shape or texture differs across the top pair and then they have to decide whether the bottom pair also has the same difference. In the control matching condition, participants are shown two objects at the top of the screen and one at the bottom, and a word in the middle of the screen (either “shape” or “texture”). They are told to decide whether the bottom object matches either of the top two objects on that dimension (e.g., if the word is “shape”, is the bottom object the same shape as either of the top two objects. For both conditions, the subject responds with a button press. For the relational condition, the stimuli are presented for 3500 ms, with a 500 ms inter-task interval, with four trials per block. In the matching condition, stimuli are presented for 2800 ms, with a 400 ms inter-task interval, and there are 5 trials per block. Each type of block (relational or matching) lasts a total of 18 seconds. In each of the two runs of this task, there are three relational blocks, three matching blocks, and three 16-second fixation blocks. In total, the relational processing task fMRI had a run duration of 2:56 minutes per run, with 232 frames per run.
            
            \item \textbf{SOCIAL:} Subjects were shown 20 second video clips of objects (squares, circles, triangles) that either interacted in some way, or moved randomly on the screen. These videos were developed by either Castelli et al. \cite{castelli2000movement}\cite{castelli2002autism} or Martin et al. \cite{wheatley2007understanding}\cite{white2011developing}. After each video clip, subjects were asked to judge whether the objects had a mental interaction (an interaction that appears as if the shapes are taking into account each other’s feelings and thoughts), Not Sure, or No interaction (i.e., there is no obvious interaction between the shapes and the movement appears random). Each of the two task runs has 5 video blocks (2 Mental and 3 Random in one run, 3 Mental and 2 Random in the other run) and 5 fixation blocks (15 seconds each). In total, the social cognition task fMRI had a run duration of 3:27 minutes per run, with 274 frames per run.
            
            \item \textbf{WORKING MEMORY:} The working memory task has been adapted from the one developed by Drobyshevsky et al., 2006 \cite{drobyshevsky2006rapid} and Caceres et al., 2009 \cite{caceres2009measuring}. Category specific representation task \cite{downing2001cortical}\cite{peelen2005within}\cite{taylor2007functional}\cite{fox2009defining} and working memory (working memory) task \cite{downing2001cortical}\cite{fox2009defining}\cite{kung2007region}\cite{peelen2005within} were combined into a single task paradigm. Subjects were presented with blocks of trials consisting of pictures of places, tools, faces, and non-mutilated body parts. Within each run, the four different stimulus types were presented in separate blocks.  Also, within each run, half of the blocks use a 2-back working memory task and the other half use a 0-back working memory task (as a working memory comparison). A 2.5 second cue indicates the task type (and target for 0-back) at the start of the block. Each of the two runs contains 8 task blocks (10 trials of 2.5 seconds each, for 25 seconds) and 4 fixation blocks (15 seconds). On each trial, the stimulus is presented for 2 seconds, followed by a 500 ms inter-task interval (ITI). In total, the working memory task fMRI had a run duration of 5:01 minutes per run, with 405 frames per run.
            
        \end{itemize}
            
    \subsection{HCP-YA preprocessing: FC pipeline}
    
        \subsubsection{The HCP-YA minimal processing pipeline overview}

            Our starting point to process the HCP-YA data is the denominated \textit{minimally processed} dataset, as provided by HCP \cite{glasser2013minimal}. The pipeline includes artifact removal, motion correction, and registration to standard space. The main steps of this pipeline are spatial (\textit{minimal}) preprocessing, in standard volumetric and combined volume and surface spaces. By taking care of the necessary spatial preprocessing once in a standardized fashion, rather than expecting each user to repeat this processing, the minimal preprocessing pipeline avoids duplicate effort and ensures a minimum standard of data quality. The main steps of this minimal processing functional pipeline \cite{smith2013resting}\cite{glasser2013minimal}\cite{smith2013resting} are described in this section.\par
            
            In total, there are six minimal preprocessing pipelines included in the HCP, three structural (\textit{PreFreeSurfer}, \textit{FreeSurfer}, and \textit{PostFreeSurfer}), two functional (\textit{fMRIVolume} and \textit{fMRISurface}), and a \textit{Diffusion Preprocessing} (not covered in this work) pipeline. Following is a brief description of the structural pipelines:
            
            \begin{enumerate}
            
                \item \textit{\textbf{PreFreeSurfer:}} This produces an undistorted "native" structural volume space for each subjects, aligns the T1w and T2w images, performs a bias field correction, and registers the subject's native structural volume space to MNI space.
                
                \item \textit{\textbf{FreeSurfer:}} This pipeline is based on FreeSurfer version 5.2. It segments the volume into predefined structures, reconstructs white and pial cortical surfaces, and performs FreeSurfer's standard folding-based surface registration to their surface atlas (\texttt{fsaverage}).
                
                \item \textit{\textbf{PostFreeSurfer:}} This pipeline produces all of the NIFTI volume and GIFTI surface files necessary for viewing the data in Connectome Workbench, applies the surface registration to the Conte69 surface template \cite{van2012parcellations}, downsamples registered surfaces for connectivity analysis, and creates the final brain mask and myelin maps.
                
            \end{enumerate}
            
            There are two volume spaces and three surface spaces in the HCP-YA data. The volume spaces are the subject's undistorted native volume space and the standard MNI space, which is useful for comparisons across subjects and studies. The surface spaces are the native surface mesh for each individual (~136k vertices, most accurate for volume to surface mapping), the high resolution Conte69 registered standard mesh (~164k vertices, appropriate for cross-subject analysis of high resolution data like myelin maps), and the low resolution Conte69 registered standard mesh (~32k vertices, appropriate for cross-subject analysis of low resolution data like fMRI or diffusion). The 91,282 standard grayordinate (CIFTI) space is made up of a standard subcortical segmentation in 2 mm MNI space and the 32k Conte69 mesh of both hemispheres. The functional and diffusion pipelines can be run after completing the structural pipelines described above. Following is a brief description of the two functional pipelines:
        
            \begin{enumerate}
            
                \item \textit{\textbf{fMRIVolume:}} This pipeline removes the spatial distortions, carries out motion correction by realigning volumes, reduces the bias field, normalizes the 4-dimensional image to a global mean, and masks the data with the final brain mask. There is no overt volume smoothing in this pipeline as the output of this pipeline is in the volume space and can be used for volume-based fMRI analysis.
                
                \item \textit{\textbf{fMRISurface:}} In this pipeline, the volume-based time series is brought into the CIFTI grayordinate standard space. The voxels within the cortical gray matter ribbon are mapped onto the native cortical surface. This transforms the voxels according to the surface registration onto the 32k Conte69 mesh and maps the set of subcortical gray matter voxels from each subcortical region in each subject to a standard set of voxels in each atlas parcel. This gives a standard set of grayordinates in every subject with 2 mm average surface vertex and subcortical volume spacing. This data is then smoothed with surface and parcel constrained smoothing of 2 mm FWHM (full width at half maximum) to regularize the mapping. This pipeline outputs a CIFTI dense time-series (denominated \texttt{\{TASK\}\_\{ACQ\}\_Atlas\_MSMAll.dtseries.nii}, where \texttt{\{TASK\}} refers to the fMRI condition and \texttt{\{ACQ\}} is the acquisition, either LR or RL) that can be used for surface-based fMRI analysis.
                
            \end{enumerate}
            
            For the resting-state data, in addition to the minimal processing pipeline described above, a 24-parameter motion regression and ICA-FIX \cite{smith2007localizing}\cite{salimi2014automatic}\cite{griffanti2014ica} have also been applied in the data provided by HCP-YA. The 24 parameters included in the motion regression step are the 6 rigid-body parameter timeseries, their backwards-looking temporal derivatives, and all squared 12 resulting regressors. The motion regressesion and ICA-FIX step applied on resting state fMRI data has produced timeseries denominated \texttt{rfMRI\_REST\_\{ACQ\}\_Atlas\_hp2000\_clean.dtseries.nii}.
            
        \subsubsection{Additional processing steps} \label{sec_add_processing}
        
            We perform the following additional steps on the fMRI data (denominated \texttt{\{TASK\}\_\{ACQ\}\_Atlas\_MSMAll\_ hp2000\_clean.dtseries.nii} for resting state and \texttt{\{TASK\}\_\{ACQ\}\_Atlas\_MSMAll.dtseries.nii} for task-based fMRI in HCP-YA data):
            
            \begin{itemize}
            
                \item \textbf{\textit{Nuisance regression:}} This step is carried out for the task fMRI data only, where we regress out the 24-parameter motion regressors (6 rigid-body parameter timeseries, their backward-looking derivatives, and all squared resulting regressors), average time-series from the cerebro-spinal fluid (CSF), and the average time-series from the white matter. We have also provided processed data where this step is not performed.
                
                \item \textbf{\textit{Global signal regression (GSR):}} This step involves the removal of the global (or average) signal from the time series of each voxel using linear regression \cite{murphy2017towards}. We have produced two sets of connectomes, one where GSR has been performed and one where it has not been, as there is a lack of consensus in the scientific community regarding whether it should be performed or not \cite{liu2017global}\cite{hayasaka2013functional}\cite{xu2018impact}\cite{gotts2013perils}\cite{saad2012trouble}\cite{aquino2020identifying}. The results shown in the main text are based on connectomes where GSR has been performed as part of the preprocessing. This step was performed for all fMRI conditions (if specified).
                
                \item \textbf{\textit{Bandpass filtering:}} Lastly, we bandpass filter the time-series data using the following parameters:
            
                \begin{itemize}
                    \item Minimum frequency ($f_{min}$) = 0.009 Hz
                    \item Maximum frequency ($f_{max}$) = 0.08 Hz for resting state and 0.25 Hz for task-based fMRI
                    \item Repetition time (\textit{TR}) = 0.72 s
                    \item Butterworth filter order = 4
                \end{itemize}
                
                This step was always performed on all the fMRI conditions.
                
            \end{itemize}
            
        \subsubsection{Brain atlases}
        
            The brain atlases used in this paper have been developed by Schaefer et al., 2018 \cite{schaefer2018local}. The Schaefer parcellations are further divisions of the resting state functional networks described by Yeo et al., 2011 \cite{yeo2011organization} and have different levels of granularity (100 to 1,000 brain regions, in steps of 100). We have also added the 14 subcortical regions as provided in the HCP-YA dataset (denominated \texttt{Atlas\_ROIs.2.nii.gz}) to each of these atlases, thus making them 114, 214,..., 1,014 brain regions. All the results described in this paper correspond to the Schaefer parcellations. 
            
            \begin{figure}[h!]
                \centering
                \includegraphics[scale=0.3]{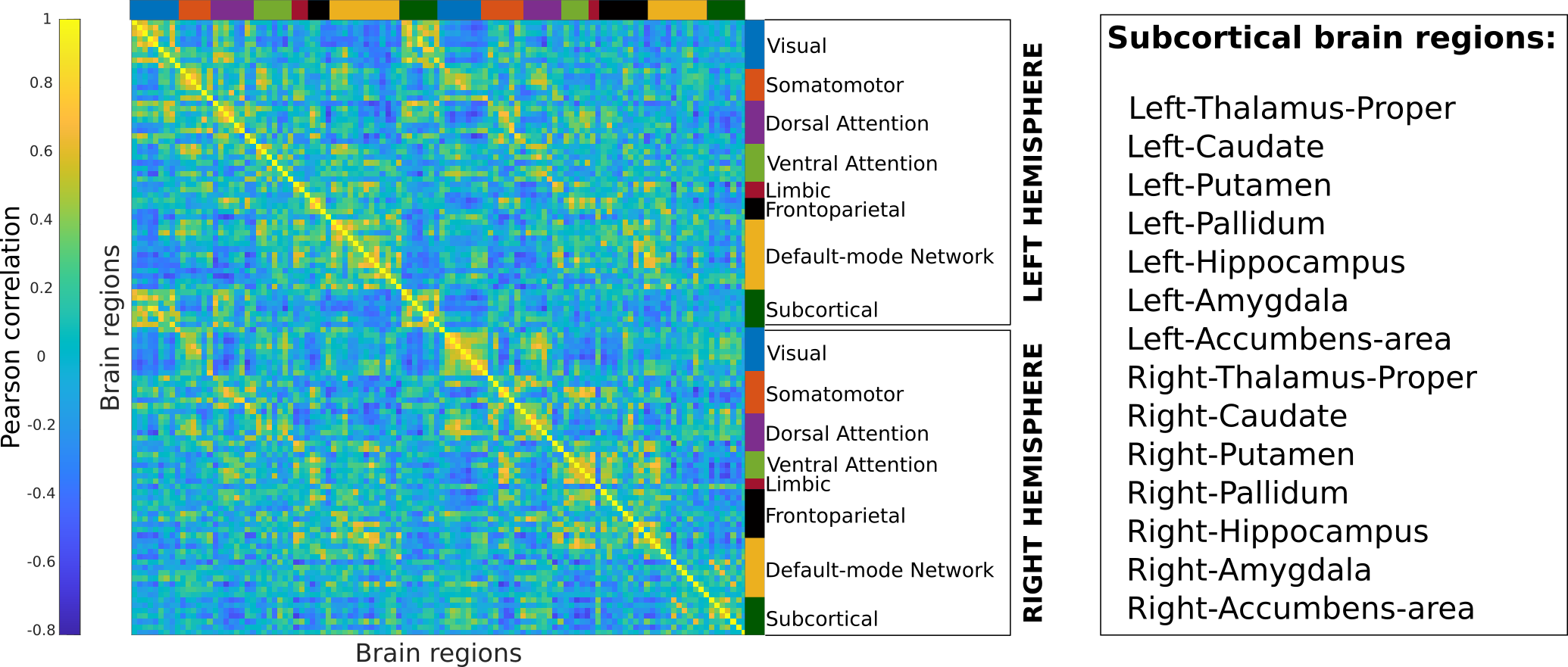}
                \caption{Example of a single-session, single-subject, whole-brain functional connectome (FC) using the Schaefer100 cortical atlas together with 14 subcortical regions. Functional couplings between brain regions are estimated through Pearson's correlation coefficients between their corresponding BOLD time-series. Rows and columns of the FC are ordered by hemisphere (Left and Right), and further divided into resting-state functional networks denoted by different colors.}
                \label{fig_sample_connectome}
            \end{figure}
            
        \subsubsection{Estimation of functional connectomes}
        
            The next step is to extract the functional time series corresponding to each brain region of the parcellation, \textit{z}-score them, and ultimately estimate the functional connectomes. This step is conducted in Connectome Workbench (freely available at \url{https://www.humanconnectome.org/software/connectome-workbench}) with the commands \texttt{cifti-reduce}, \texttt{cifti-math}, and \texttt{cifti-parcellate}. The command \texttt{cifti-reduce} is used to compute the mean and standard deviation of time series data, while \texttt{cifti-math} computes the \textit{z}-scored time series data using the mean and standard deviation previously computed. Lastly, \texttt{cifti-parcellate} parcellates the voxel-wise time series data into different brain regions as per the Schaefer parcellation, by averaging all the voxel-level time series belonging to each brain region. These parcellated time series have been made available and can be used to perform brain-region level activity and connectivity analysis. \par
            
            Finally, whole-brain functional connectomes are generated by computing the Pearson's correlation coefficients between the time series of every pair of brain regions in the parcellated time series data computed in the earlier step. These connectomes are square and symmetric matrices that have been made available and can be used to perform functional connectome analyses. \textit{Figure \ref{fig_sample_connectome}} shows a sample connectome parcellated into the Schaefer100 atlas, with 114 brain regions (100 cortical + 14 subcortical). The figure names all the subcortical regions.
    
    \subsection{The differential identifiability framework ($\mathbb{I}\mathit{f}$)}
    
        \subsubsection{Identifiability matrix}
        
            In order to quantify the subject-level fingerprint from a cohort of functional connectomes, Amico and Goñi \cite{amico2018quest} proposed an object called the \textit{identifiability matrix}. This is a non-symmetric correlation matrix that compares the all-to-all test-retest functional connectomes from a cohort of unrelated subjects. Thus, every entry in this matrix is the Pearson's correlation between test and retest functional connectomes in their vector form. Typically, the \textit{x}-axis of the identifiability matrix represents the test (or first run, or day 1) and \textit{y}-axis the retest (or second run, or day 2). Importantly, the ordering of the subjects is kept the same in the rows and columns of this matrix (i.e., test and retest sessions), and hence the main diagonal contains the correlation values between the test and retest connectomes of the same subject. The higher the values in the main diagonal compared to the off-diagonal elements, the better the subject-level fingerprint of the dataset. \par
            
\begin{figure}[h!]
                \centering
                \includegraphics[scale=0.27]{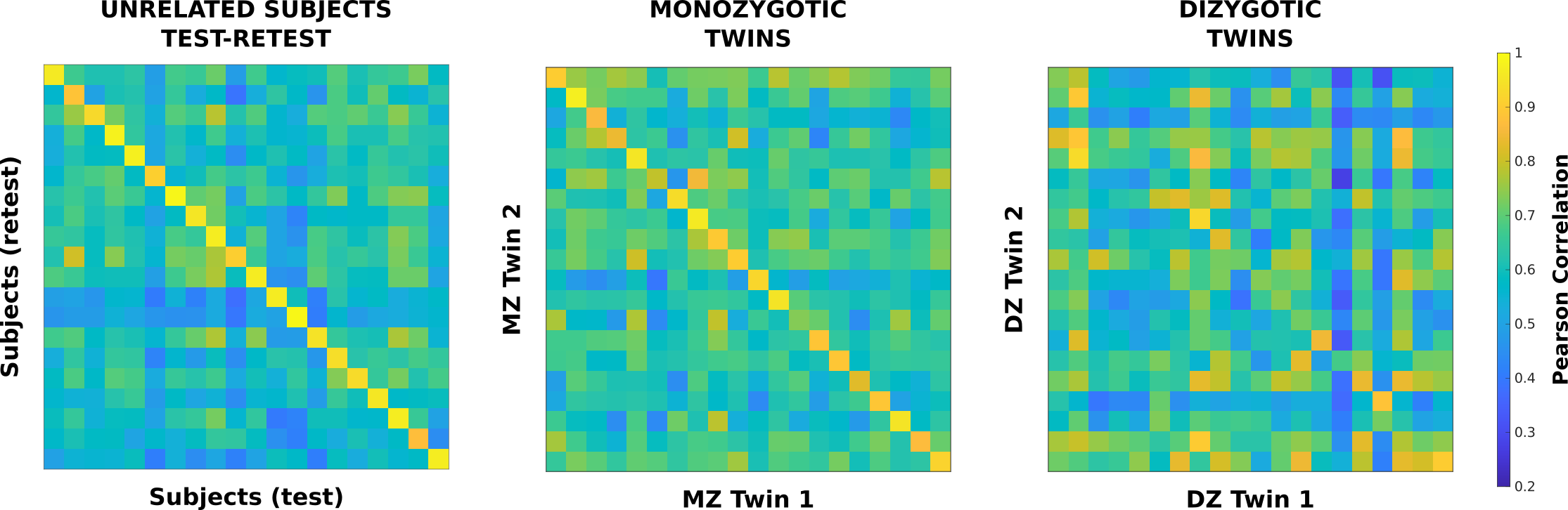}
                \caption{Differential identifiability matrices for sample cohorts of 20 Unrelated subject test-retest, Monozygotic twin pairs, and Dizygotic twin pairs}
                \label{fig_identmat}
            \end{figure}

            We have further expanded on this intuitive interpretation of the identifiability matrix by using the connectomes from twin pairs instead of test-retest of the same subject. In order to do so, we are extending the concept of identifiability and fingerprints beyond test/retest of the same subjects. To do so, we have computed the identifiability matrix for two different cohorts of twins, monozygotic (MZ) twins and dizygotic (DZ) twins. In this case, rows and columns of the identifiability matrix represent each of the two twins respectively, with the main diagonal values being the Pearson correlations between the FCs of the twin pair and the off-diagonal being the correlations between the FCs of unrelated subjects from the twin-cohort. \textit{Figure \ref{fig_identmat}} shows the differential identifiability matrices for sample cohorts of 20 test-retest, MZ twin pairs, and DZ twin pairs. \par
            
        \subsubsection{Differential identifiability score}
        
            Using the identifiability matrix, Amico and Goñi \cite{amico2018quest} have proposed a measure called \textit{differential identifiability} ($I_{diff}$) to quantify the subject-level fingerprint. $I_{diff}$ quantifies the contrast between the self-similarity (main diagonal) and the similarity between different subjects (off diagonal). $I_{diff}$ can be computed as:
            
            \begin{equation}
            \label{eqn_idiff_self}
                I_{diff}^{subject} = (I_{self} - I_{others}) \times 100
            \end{equation}
            
            \noindent where, \par
            
            \noindent $I_{self}$ = self similarity, mean of the main diagonal values in the identifiability matrix \par
            
            \noindent $I_{others}$ = similarity between different subjects, mean of the off-diagonal elements in the identifiability matrix \par
            
            As discussed above, the differential identifiability score can also be calculated for a twin cohort by pairing FCs of twin subjects instead of test-retest. In this case, the main diagonal elements of the identifiability matrix will be the correlations between the FCs of twin subjects (MZ and DZ) and the off-diagonal elements will be the correlations between unrelated subjects. In this case, the \textit{twins} differential identifiability can be expressed as:
        
            \begin{equation}
            \label{eqn_idiff_twin}
                I_{diff}^{twin} = (I_{twin} - I_{others}) \times 100
            \end{equation}
            
            This can be repeated separately for monozygotic (or identical) twins and dizygotic (or fraternal) twins. Monozygotic (MZ) twins are genetically 100\% identical whereas dizygotic (DZ) twins have, on average, 50\% genetic material in common \cite{prescott1995twin}\cite{blokland2013twin}.
            
        \subsubsection{PCA-based differential identifiability framework}
        
            In order to assess and compare the different Schaefer parcellations with each other in terms of their fingerprints, we have adapted the identifiability framework put forth by Amico and Goñi, 2018 \cite{amico2018quest}. They used group-level principal component analysis (PCA) to decompose functional connectomes into orthogonal \textit{principal components} and then subsequently reconstructed with fewer and fewer principal components in order to find the reconstruction level where the differential identifiability score was maximum. Further developments based on the differential identifiability framework have been recently used to improve FC fingerprints across different scanning sites \cite{bari2019uncovering} as well as in network-derived measurements \cite{rajapandian2020uncovering}. PCA is a statistical procedure that transforms a set of observations of possibly correlated variables into a set of linearly uncorrelated variables, i.e., \textit{principal components}. PCA as a tool is widely used in the exploratory analysis of the underlying structure of data in pattern recognition \cite{deng2013nonlinear}\cite{hsieh2009novel} and denoising \cite{manjon2013diffusion}\cite{de2007denoising}, among other areas.  \par
            
            \begin{figure}[h]
                \centering
                \includegraphics[scale=0.14]{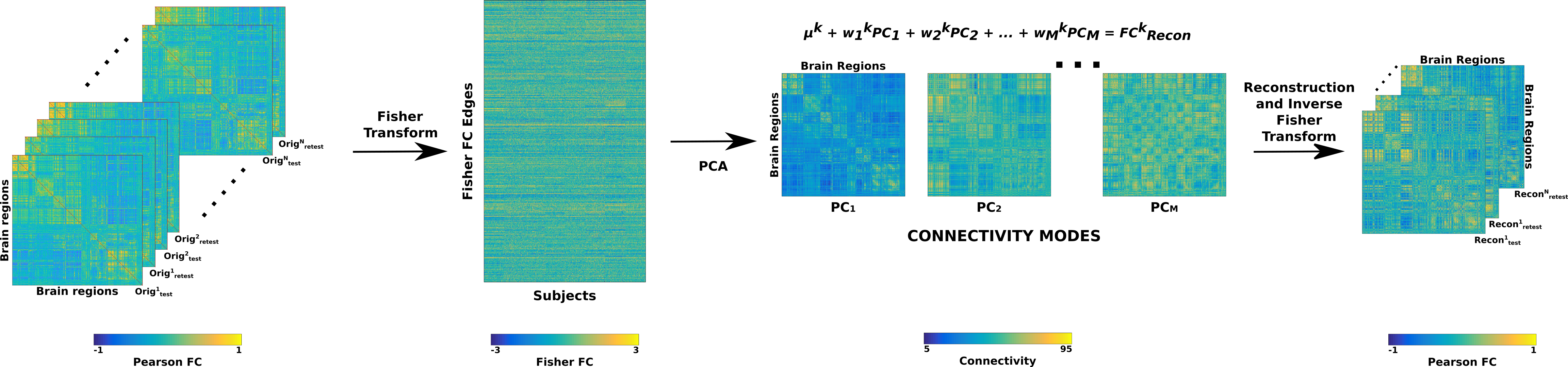}
                \caption{Workflow scheme of the group-level principal component analysis (PCA) reconstruction procedure of individual functional connectomes. The upper triangular values (as the matrices are symmetrical) of the test and retest FCs are vectorized, \textit{z}-transformed using Fisher transform (MATLAB function \texttt{atanh}), and stacked into a matrix. This matrix is then decomposed using PCA to get as many components as connectomes in the cohort. The next step is to incrementally add principal components to the reconstruction, undo the Fisher transform (MATLAB function \texttt{tanh}) to get reconstructed functional connectomes, and compute the differential identifiability at each step. }
                \label{fig_PCA}
            \end{figure}
            
            We noticed that during the partial reconstructions of the functional connectomes using subsets of principal components, the FCs were not pure correlation matrices as some of the values in the FCs fell outside the [-1 1] range. To avoid this numerical issue, we have adapted the identifiability framework as proposed by Amico and Goñi for this study by using Fisher transform \cite{fisher1915frequency} as shown in \textit{Figure \ref{fig_PCA}}. As can be seen in the figure, we vectorize upper triangular (as the matrices are symmetrical) values in each FC (two FCs per subject for test-retest and one FC per subject for twins) before assembling them into a matrix where the columns are separate FCs and rows are the vectorized connectivity patterns. Before this vectorization, we \textit{z}-transform the FCs by employing Fisher transform (MATLAB function \texttt{atanh}). Fisher transform has also been employed previously in several studies focusing on functional connectivity in the human brain \cite{negishi2011functional}\cite{hampson2010functional}\cite{tomasi2012resting}\cite{fox2013identification}. The assembled matrix of Fisher transformed FCs is then decomposed using PCA into as many principal components as input FCs. In the next step, we reconstruct the FCs by incrementally adding one principal component at a time (in descending order of explained variance) and employing the inverse Fisher transform (MATLAB function \texttt{tanh}) in order to get back the Pearson correlation-based FCs. These FCs at each step of the reconstruction are then used to compute the identifiability matrix and, by extension, the differential identifiability score ($I_{diff}$). Through this procedure, we obtain a curve of $I_{diff}$ values for the whole range of principal components used in the reconstruction. It should also be noted that when all the principal components are used to reconstruct the FCs, we obtain the original input FCs, and thus the resulting $I_{diff}$ score corresponds to that of the original FCs.
            
    \subsection{Assessment of brain fingerprints}
    
        Using the identifiability framework, we have assessed the brain fingerprints for three different cohorts -- test-retest of a group of unrelated subjects (for the rest of the paper, we refer to this as \textit{Unrelated subjects}), MZ twin pairs, and DZ twins pairs -- at all the different levels of granularity afforded to us by the Schaefer parcellations, and for all fMRI conditions. We have set up different experimental designs to evaluate the fingerprints at the whole brain level and resting state functional network level. We also examine the effect of scanning length and repetition time (TR) on the subject-level and twin fingerprints for resting state fMRI. This section contains the description of the different experimental designs. \par
        
        For consistency, we have only included those subjects in each of these three datasets in the PCA identifiability framework (unrelated subjects, MZ twins, and DZ twins) for whom all the task test/retest (both runs) functional connectomes are available. Thus, there are 428 unrelated subjects, 116 MZ twins pairs, and 63 DZ twin pairs included in the PCA reconstruction.
            
        \subsubsection{Whole brain differential identifiability}
        
            For a given parcellation granularity and a given fMRI condition, whole-brain FCs are used to compute the differential identifiability profiles.
             
        \subsubsection{Comparison of individual and twin fingerprint}
        
            To facilitate meaningful comparison between the differential identifiability profiles between the Unrelated subjects, MZ twins, and DZ twins, we have run the identifiability framework on a subset of Unrelated subjects and MZ twin data so that the number of FCs in these cohorts is equal to the number of FCs in the DZ twin dataset (as DZ is the smallest dataset). This analysis was performed only for the Schaefer400 parcellation for all fMRI conditions. We have conducted the analysis for 100 bootstrap runs of 80\% connectome pairs for each of the three cohorts. We hypothesize an ordinal presence of fingerprints that is highest for test-retest, lower for MZ twins, and lowest for DZ twins. \par
            
            For each cohort separately, we test a null model where the rows of the identifiability matrix are shuffled before computing the identifiability score. We perform the bootstrap runs for the null models as well. This is to test whether the identifiability values we obtain from the identifiability framework applied to the three cohorts are a matter of chance.
            
        \subsubsection{Functional network-specific differential identifiability}
        
            In order to quantify the amount of fingerprint specific to a functional network, we assess the differential identifiability profiles by considering only the brain regions inside a specific functional network. For the network definitions, we use the 7 resting-state networks (RSNs) provided by Yeo et al., 2011 \cite{yeo2011organization}. We should highlight that the PCA decomposition for this experiment is identical to when we explore whole-brain differential identifiability as described above, but for the differential identifiability calculation we only include the brain regions that belong to a specific RSN.
            
        \subsubsection{Effect of scanning length on differential identifiability} \label{sec_scan_len}
        
            In order to test the effect of scanning length (in term of number of frames) on differential identifiability, we compute the differential identifiability profiles by gradually increasing the sequential number of fMRI volumes used in constructing the FCs. For this analysis, we have used resting-state scans as they have the longest scan duration (approx. 15 minutes) and the highest number of frames (1200). This allows us to study the effect of scanning length on differential identifiability for a wide range (50 to 1200, in steps of 50). 
            
\section{Results} \label{sec_results}

    \subsection{The HCP-YA Functional Connectomes Data Release}
    
\begin{figure}[h!]
            \centering
            \includegraphics[scale=0.4]{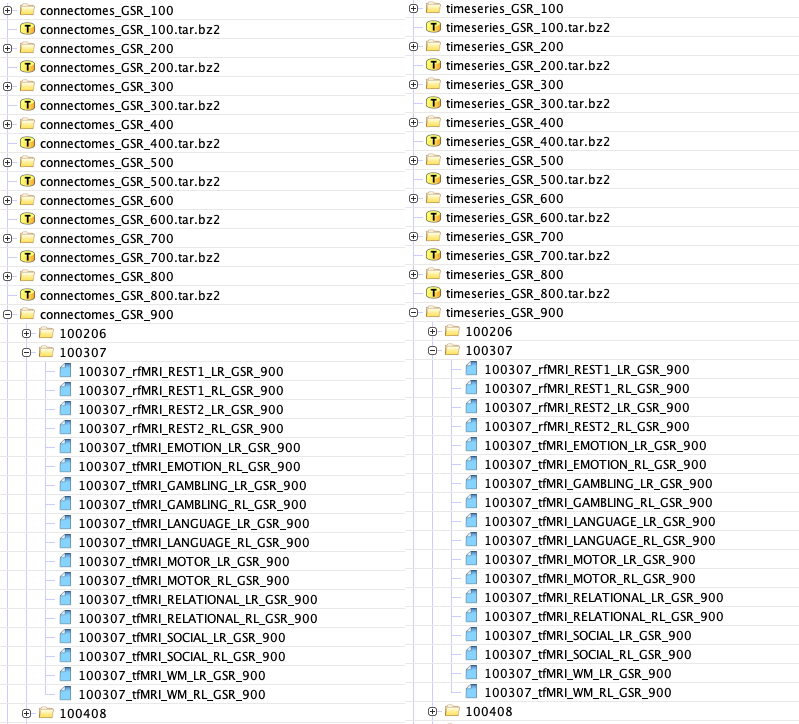}
            \caption{Sample data structure for functional connectomes and parcellated timeseries}
            \label{fig_data_structure}
        \end{figure}

        The results of the processing performed in \textit{Section \ref{sec_methods}} have been made publicly available at \censor{\url{https://rdl-share.ucsd.edu/message/0Y3GKJM7a2CR2FgMSbk4st}} (functional connectomes) and \censor{\url{https://rdl-share.ucsd.edu/message/Lqi0Oj0fALIrh4lv0Z5L06}} (parcellated time-series). Here, we outline the specific data products that have been produced and how to access them. The data release includes FCs and time-series parcellated according to the Schaefer atlases \cite{schaefer2018local} with different levels of granularity. For ease of downloading, we have created separate compressed files for each of the Schaefer parcellations, GSR/non-GSR status, and connectome/timeseries data. Additionally, for the task-based fMRI, we also have FCs with and without 26 regressors that correspond to motion and average signals from white matter and CSF (see details in \textit{Section \ref{sec_add_processing}}). Each compressed file includes the FCs or time-series (depending on the selection) of all the fMRI conditions (resting state and 7 tasks) included in the HCP dataset that we have parcellated into the selected granularity of Schaefer parcellation. \textit{Figure \ref{fig_data_structure}} shows an example of the data structure for connectome and timeseries data, respectively, for data processed with GSR. \par
        
    \subsection{Whole brain differential identifiability}
    
        \subsubsection{Unrelated subjects test-retest}
        
\begin{landscape}
                \begin{figure}[htbp!]
                    \centering
                    \includegraphics[scale=0.15]{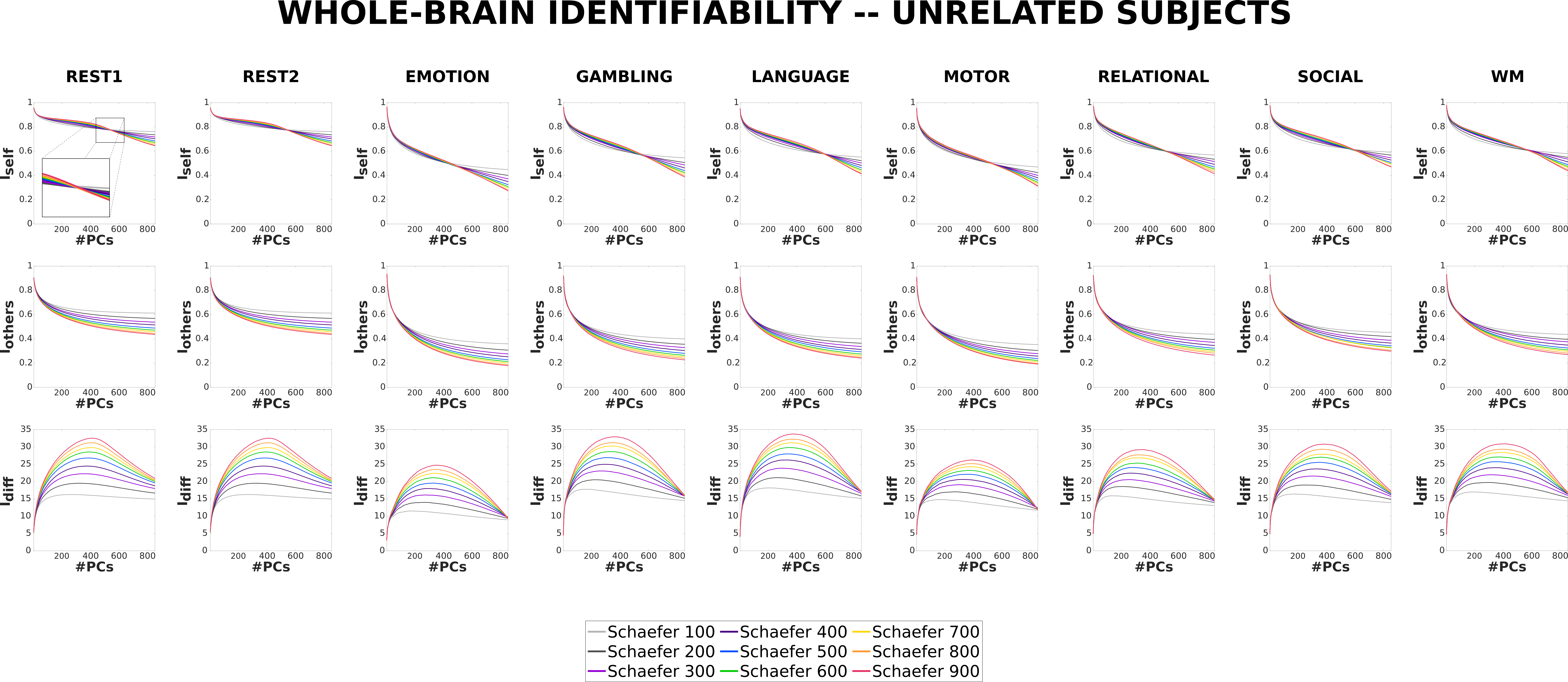}
                    \caption{$I_{self}$, $I_{others}$, and $I_{diff}$ curves for Schaefer 100 to 900 parcellations for all the fMRI conditions in HCP, for unrelated subjects. The higher the granularity of the Schaefer parcellation, the higher the test-retest identifiability regardless of the fMRI condition.}
                    \label{fig_unrelated}
                \end{figure}
            \end{landscape}

            We first ran and assessed the differential identifiability framework on Unrelated subjects. This assessment is an extension with respect to Amico and Goñi \cite{amico2018quest} where a small cohort was evaluated (100 unrelated subjects, as opposed to 428) by using a single parcellation scheme \cite{glasser2016multi}. \textit{Figure \ref{fig_unrelated}} shows the $I_{self}$, $I_{others}$, and $I_{diff}$ profiles for all the Schaefer parcellations for the 7 tasks and resting state included in the HCP-YA dataset. \par
            
            Please note that in all the plots ($I_{self}$, $I_{others}$, and $I_{diff}$), the values for the maximum number of principal components correspond to the full reconstruction of the FCs with all the variance retained, i.e., original FCs. The number of principal components for which $I_{diff}$ is highest is considered the optimal point of reconstruction. As can be seen, the $I_{self}$ value decreases with increasing granularity for original FCs. However, as we reconstruct with fewer principal components, the $I_{self}$ values pass through a point of inflection (where $I_{self}$ values are approximately equal for all granularities) and leads to the optimal point of reconstruction where the reverse is true; i.e. $I_{self}$ increases with increasing granularity. $I_{others}$ on the other hand, is consistently lower for higher granularity across the whole range of principal components. Finally, $I_{diff}$ values for original FCs are either approximately equal (e.g.n in MOTION) or increase negligibly with increasing granularity. However, at the optimal point of reconstruction, the $I_{diff}$ score always increases with increasing granularity and the difference between the $I_{diff}$ curves is much more pronounced. \par
            
        \subsubsection{Monozygotic twins}
        
            Monozygotic (or identical) twins share 100\% of their genetic material \cite{blokland2013twin}\cite{prescott1995twin}. The differences between the MZ twins thus arise from their having different environments. \par
        
            Both $I_{twins}$ and $I_{others}$ for MZ twins decrease with increasing granularity for the original FCs. $I_{diff}$ scores are approximately equal across all granularities for original FCs. However, similarly as in Unrelated subjects, $I_{diff}$ score increases with increasing granularity and the difference between the $I_{diff}$ profiles for MZ twins is prominent at the optimal reconstruction.
            
            \begin{landscape}
                \begin{figure}[htbp!]
                    \centering
                    \includegraphics[scale=0.15]{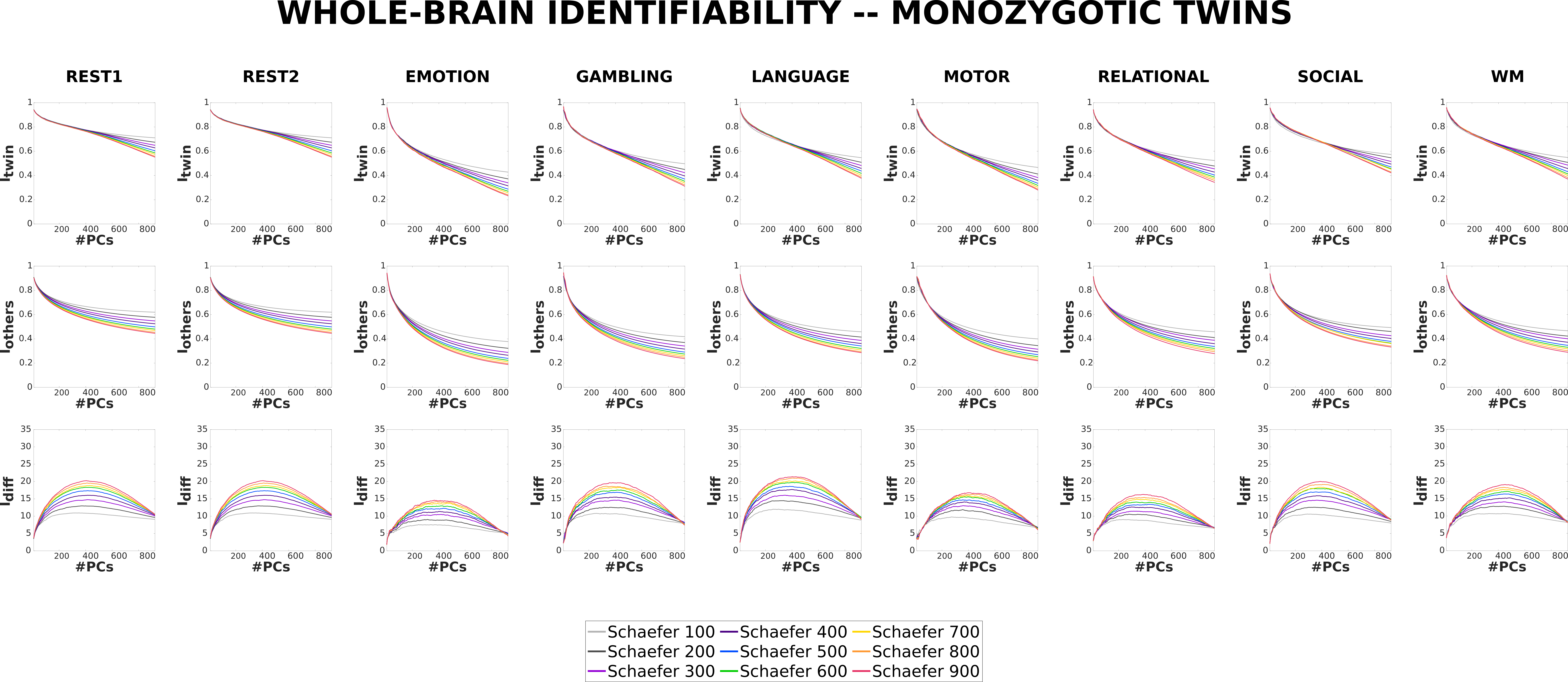}
                    \caption{$I_{self}$, $I_{others}$, and $I_{diff}$ curves for Schaefer 100 to 900 parcellations for all the fMRI conditions in HCP, for monozygotic (MZ) twin subjects. The higher the granularity of the Schaefer parcellation, the higher the MZ twin identifiability regardless of the fMRI condition, although the differential identifiability of MZ twins is lower than that of test-retest of the same subject.}
                    \label{fig_MZ_wb}
                \end{figure}
            \end{landscape}
 
        \subsubsection{Dizygotic twins}

            Dizygotic (or fraternal) twins are, on average, share 50\% of their genetic material \cite{blokland2013twin}\cite{prescott1995twin}. Thus, genetically speaking, they are siblings, but often their environment is more similar than that of non-twin siblings as they are born at the same time \cite{mark2017using}. \par
            
            Similar to the results of MZ twins, both $I_{twins}$ and $I_{others}$ for DZ twins also decrease with increasing granularity for the original FCs. $I_{diff}$ scores are approximately equal across all granularities for original FCs and, similar to Unrelated subjects and MZ twins, $I_{diff}$ score increases with increasing granularity at the optimal point of reconstruction. However, the difference between the $I_{diff}$ profiles for DZ twins is not as prominent as that for Unrelated subjects or MZ twins, but is still noticeable. 
            
            \begin{landscape}
                \begin{figure}[htbp!]
                    \centering
                    \includegraphics[scale=0.15]{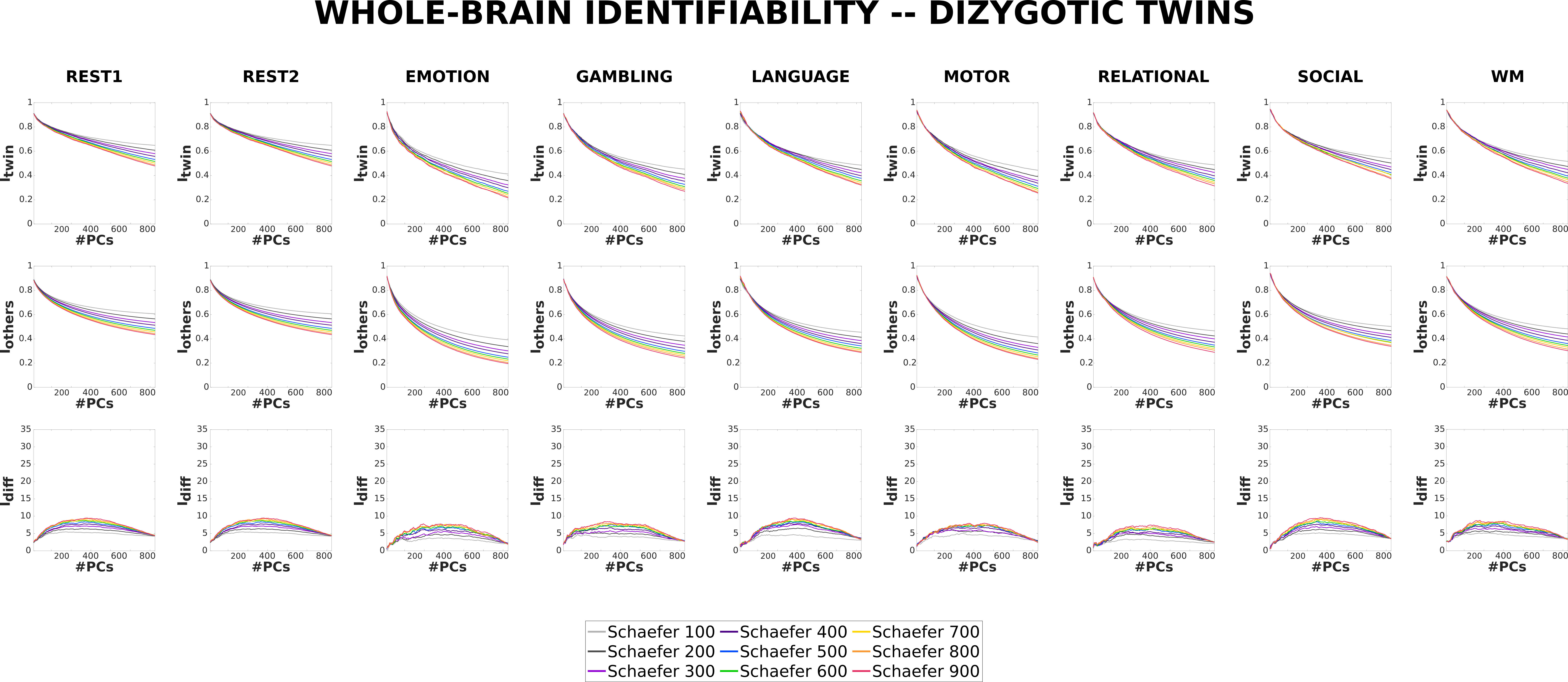}
                    \caption{$I_{self}$, $I_{others}$, and $I_{diff}$ curves for Schaefer 100 to 900 parcellations for all the fMRI conditions in HCP, for dizygotic (DZ) twin subjects. The higher the granularity of the Schaefer parcellation, the higher the DZ twin identifiability regardless of the fMRI condition. The differential identifiability of DZ twins is lower than that of both MZ twins and test-retest of the same subject.}
                    \label{fig_DZ_wb}
                \end{figure}
            \end{landscape}
            
        \subsubsection{ Comparison of individual and twin fingerprint}
    
\begin{figure}[htbp!]
                \centering
                \includegraphics[scale=0.25]{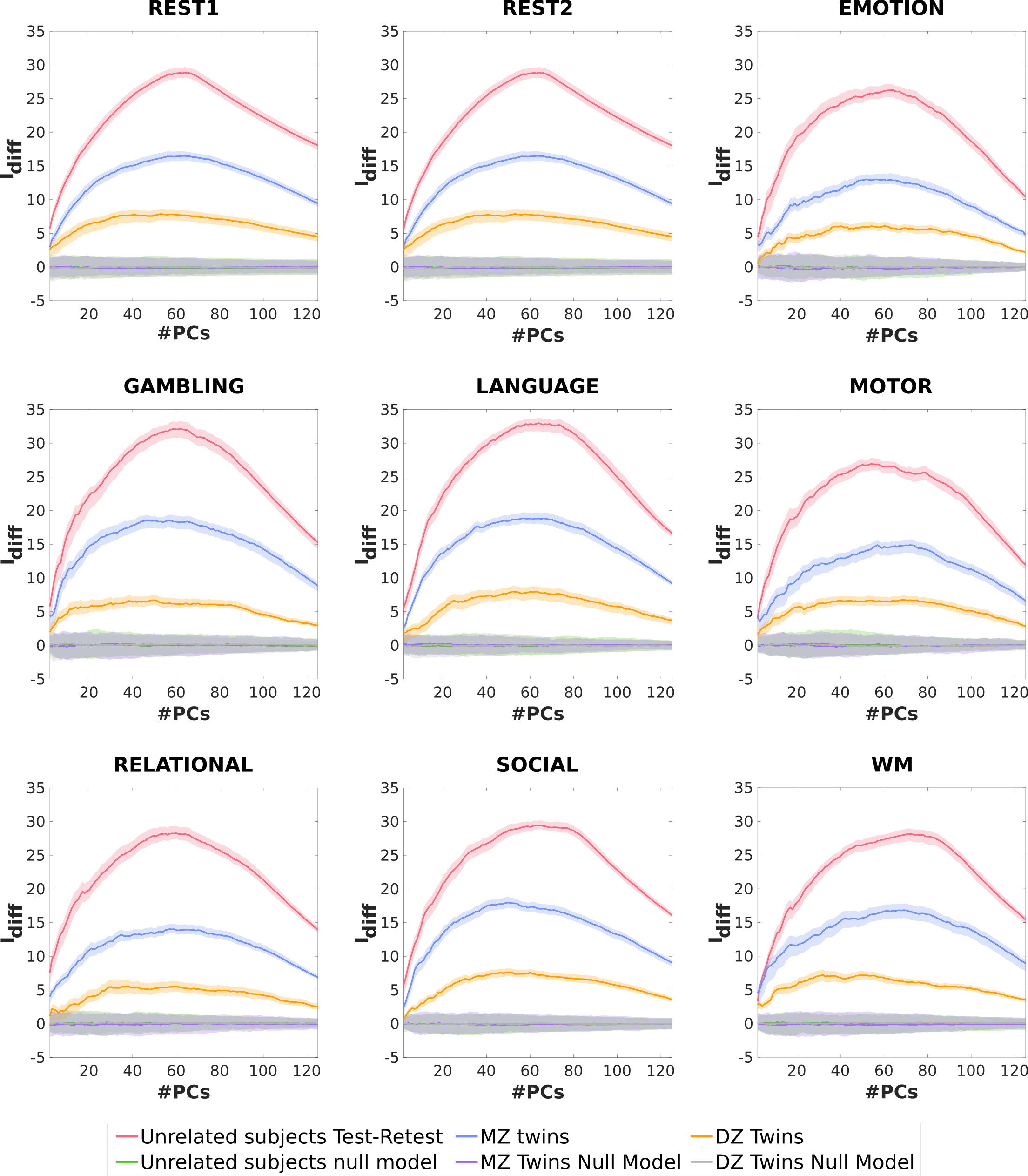}
                \caption{$I_{diff}$ profiles for the three cohorts -- Unrelated subject test-retest (red), Monozygotic twins (blue), and Dizygotic twins (orange) -- for all fMRI conditions using Schaefer400 parcellation. The cohort sizes have been matched in order to facilitate comparisons between them. The figure also includes results for the null models based on the three cohorts. Shaded areas represent the variability (5-95 percentile) of $I_{diff}$ scores across the 100 samples without replacement.}
                \label{fig_continuum}
            \end{figure}

            In order to facilitate a meaningful comparison across the three cohorts (Unrelated subjects, MZ twins, and DZ twins), we chose a random subset of Unrelated subjects and a separate random subset of MZ twins so that their numbers match the sample size of DZ twins (DZ twins cohort has the smallest sample size out of the three cohorts). In \textit{Figure \ref{fig_continuum}}, we have plotted the $I_{diff}$ profiles for Unrelated subjects, MZ twins, and DZ twins cohorts for all tasks and resting state for the Schaefer400 parcellation.
            
            As can be seen in \textit{Figure \ref{fig_continuum}}, the $I_{diff}$ scores are the highest for Unrelated subjects across all the fMRI conditions for the entire range of principal components. These are then followed by the $I_{diff}$ scores for MZ twins and DZ twins, respectively. The three curves at the bottom of the figures are results for the null models. As can be seen, the $I_{diff}$ scores for these null models are approximately zero for the entire range of the principal components (two of the curves are mostly hidden behind the third).
              
        \subsubsection{Functional network-specific differential identifiability}
    
\begin{figure}[htbp!]
                \centering
                \includegraphics[scale=0.35]{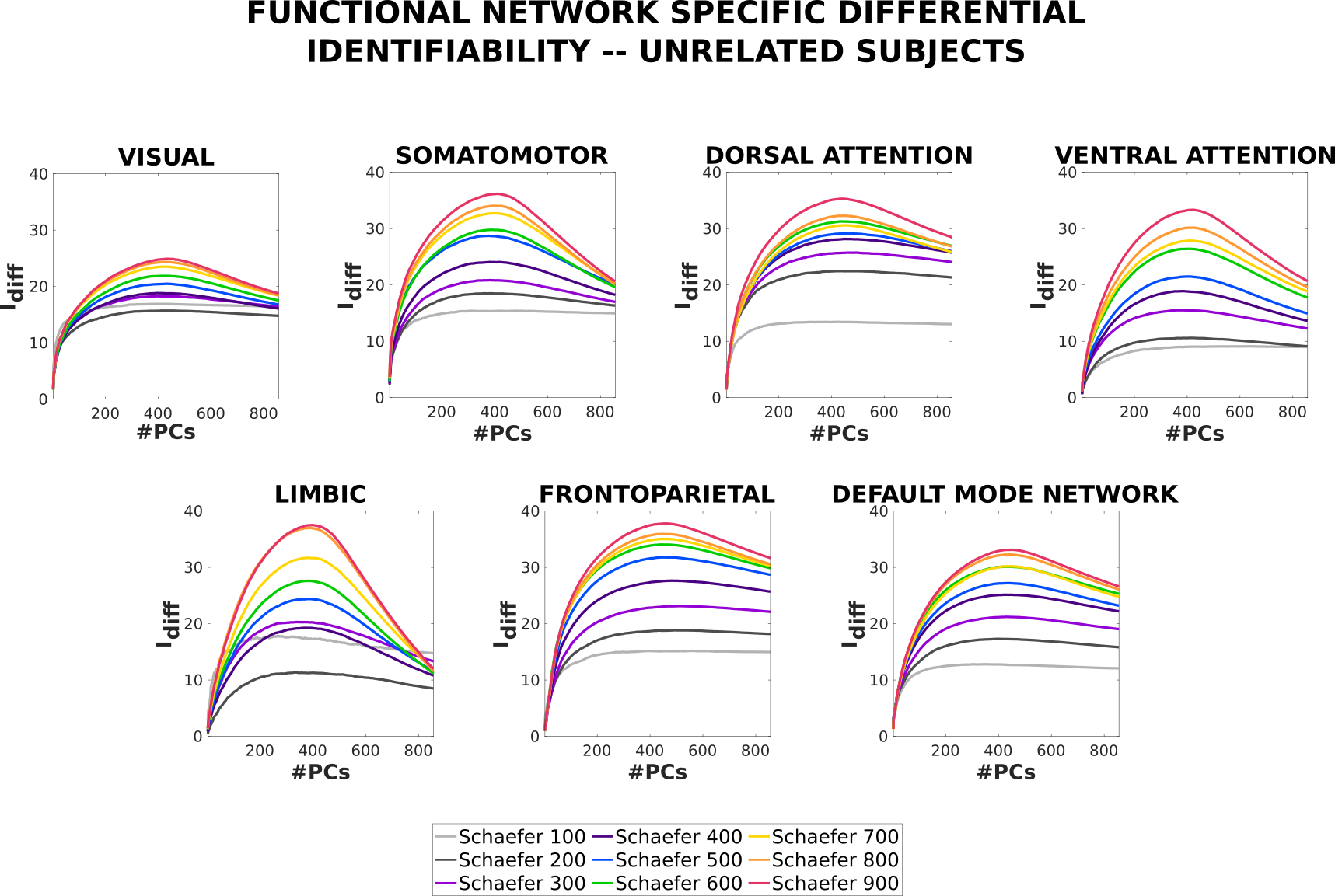}
                \caption{Functional network-specific $I_{diff}$ curves for Schaefer 100 to 900 parcellations for resting state connectomes. The higher the granularity, the higher the differential identifiability in most cases. This does not hold true when the number of brain regions included in a functional network is too few. For example, there are less than 10 brain regions included in the limbic functional network for the Schaefer 100 parcellation, which causes the $I_{diff}$ curve to be very unstable. }
                \label{fig_idiff_yeo}
            \end{figure}

            In order to assess the level of fingerprint in specific functional networks of the brain, we have computed the $I_{diff}$ profiles of each of the 7 resting-state networks, as proposed by Yeo et al. \cite{yeo2011organization}. \textit{Figure \ref{fig_idiff_yeo}} shows the differential identifiability profiles for the 7 RSNs using resting-state FCs for all Schaefer parcellations. Please note that the decomposition/reconstruction based on PCA is carried out on whole-brain FCs and not on isolated functional networks. In other words, results presented in this section belong to the same decomposition/reconstruction procedure as the ones shown in \textit{Figure \ref{fig_unrelated}}. \par
            
            When assessing $I_{diff}$ in an isolated fashion on each RSN, it can be observed that granularity of the parcellations increases $I_{diff}$ at any level of reconstruction. Notice that some RSNs present higher levels of fingerprints than others for all granularities. For any RSN (with the only exception of VISUAL), at the optimal level of reconstruction the $I_{diff}$ scores for any given Schaefer parcellation are higher than that achieved when computing the $I_{diff}$ scores using whole-brain FCs (see \textit{Figure \ref{fig_unrelated}}).
            
        \subsubsection{Effect of scanning length on differential identifiability} \label{sec_scanlen_effect}
            
\begin{figure}[htbp!]
                \centering \includegraphics[width=\textwidth]{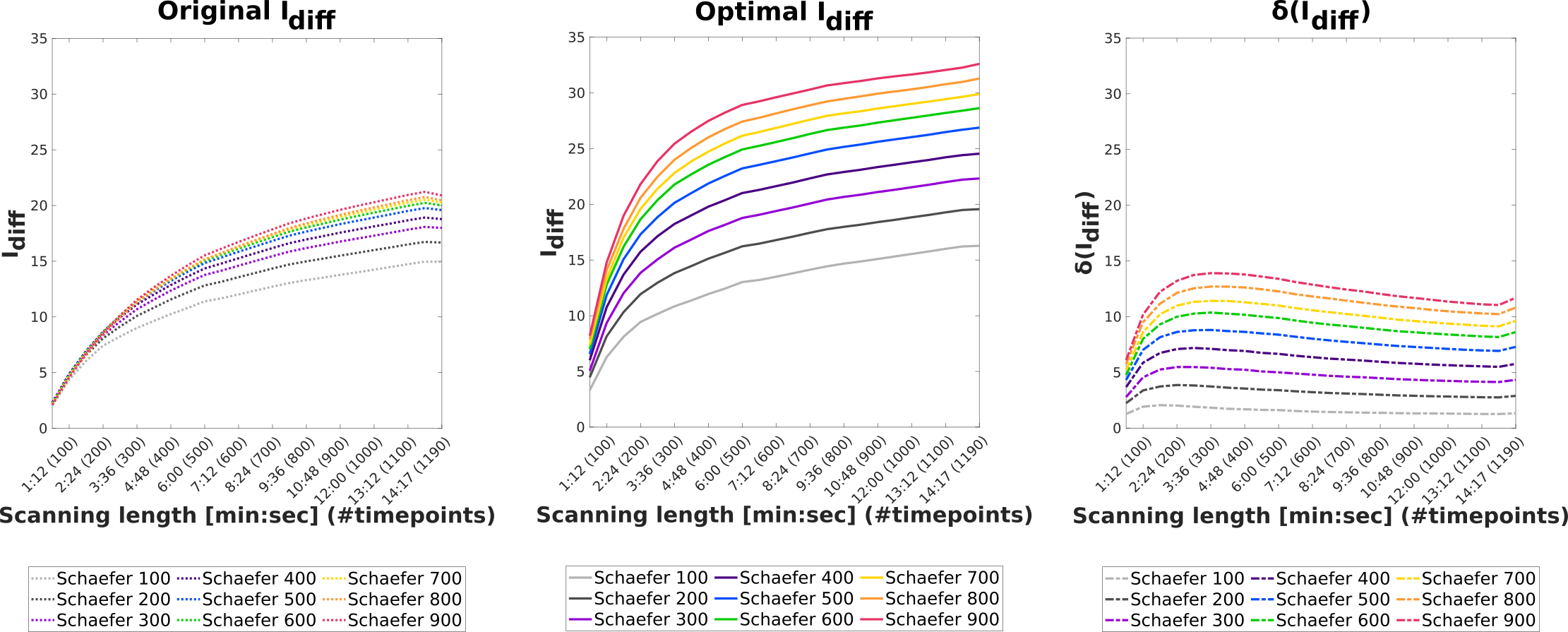}
                \caption{Original and optimal $I_{diff}$ values for resting state in all Schaefer parcellations for different scanning lengths, along with the difference between the two. For every Schaefer parcellation, we mimic a shorter scanning length by sampling from the entire rs-fMRI scan (50:50:1190 timepoints), construct functional connectomes from these shortened scanning lengths, and run the PCA identifiability framework in order to study their stability. The \textit{x}-axes of the plots show the scanning length, both in terms of minutes and seconds and the number of timepoints.  }
                \label{fig_reduced_tseries}
            \end{figure}

            For the next analysis, we assessed the effect of different lengths of acquisition time on the differential identifiability of functional connectomes. In order to do so, we subsampled different lengths of time series from the resting state fMRI acquisition and applied the identifiability framework on the resulting connectomes. We repeated this experiment for all the Schaefer parcellations. \textit{Figure \ref{fig_reduced_tseries}} shows the original and optimal differential identifiability for different number of time points and for all Schaefer parcellations, along with the difference between them. Also observe that the difference between original and optimal $I_{diff}$ increases with the granularity of the brain atlas. From the plot showing the difference between the original and optimal $I_{diff}$ scores, we can also note that, for every parcellation and scanning length combination, the differential identifiability is always higher. The different levels of granularity are more distinguishable from each other in terms of their $I_{diff}$ scores for shorter scanning lengths ($>$150 timepoints) at the optimal reconstruction as compared to the original FCs ($>$300 timepoints).
        
\section{Discussion}
    
    In this paper, we have discussed the processing pipeline we have developed to extract brain region-level time-series and the subsequent functional connectomes for each session and all the fMRI conditions of all the subjects in the Human Connectome Project -- Young Adult (HCP-YA) dataset. We have made these time-series and functional connectomes datasets available, parcellated according to the Schaefer atlases \cite{schaefer2018local} that afford us different levels of granularity (100 to 900 brain regions, in steps of 100), combined with subcortical regions (7 regions in each hemisphere; 14 in total). We have also provided a quantification of the individual and twin (monozygotic and dizygotic) fingerprint present in the FCs (of each fMRI condition separately) using an extension of the identifiability framework proposed by Amico and Goñi, 2018 \cite{amico2018quest}. Briefly, results show the presence of fingerprints at three different levels of genetic and environmental similarity (as depicted by Unrelated subjects greater than MZ twins, greater than DZ twins; \textit{Figure \ref{fig_continuum}}).These results are present for all fMRI conditions evaluated and with different sensitivity to parcellation granularity. We also found that the identifiability framework not only uncovers individual and twin-fingerprints in FCs, but, importantly, also enables us to benefit from the higher levels of fingerprints present in FCs corresponding to higher levels of granularity (\textit{Figures \ref{fig_unrelated}, \ref{fig_MZ_wb}}, and \textit{\ref{fig_DZ_wb}}). Subsequently, we discovered that different levels of fingerprint are present for the various resting-state networks, with the same pattern of higher levels of granularity enabling us to uncover higher fingerprints (\textit{Figure \ref{fig_idiff_yeo}}). Finally, we found that the amount of individual fingerprint in resting-state FCs increases with increasing scanning length, but saturating after $\sim$13 minutes of scanning (\textit{Figure \ref{fig_reduced_tseries}}).  \par
    
    We have assessed the individual-level fingerprint between the test and retest FCs of the cohort of Unrelated subjects using the identifiability framework (see \textit{Figure \ref{fig_unrelated}}). Consistent with previous investigations \cite{amico2018quest}\cite{finn2015functional}\cite{bari2019uncovering}\cite{abbas2020geff}\cite{pallares2018extracting}\cite{liu2018chronnectome}\cite{byrge2019high}\cite{mueller2013individual}\cite{faskowitz2020edge}\cite{venkatesh2020comparing}\cite{satterthwaite2018personalized}\cite{gratton2018functional}\cite{mars2018connectivity}, we have found that FCs have a recurrent and reproducible individual fingerprint across all the fMRI conditions. This is all the more important as in this study, the sample size of upwards of 400 Unrelated subjects is considerably bigger than the previous studies (usually 100 unrelated subjects from the HCP-YA dataset). As can be seen in \textit{Figure \ref{fig_unrelated}}, without implementing the identifiability framework and hence assessing original FCs, there is little to no difference in $I_{diff}$ scores between the different parcellations for each of the fMRI conditions. This implies that the granularity of the parcellation is inconsequential in terms of individual fingerprint, prompting one to use the smallest parcellation as it would lead to lower computational load. However, the difference between the fingerprints of the different parcellations only becomes apparent when the identifiability framework is applied (see \textit{Figure \ref{fig_unrelated}}). In particular, it can be observed that the higher the granularity of a parcellation, the higher the optimum test-retest fingerprint achieved for the cohort of Unrelated subjects. In other words, the potential to uncover fingerprints by fine-grained parcellations of the cortex is only unleashed when using the identifiability framework. On a related note, higher granularity was associated with a larger number of principal components leading to the highest $I_{diff}$ scores. This might be an indication of higher granularity parcellations containing more information about the individual fingerprint. \par
    
    A subset of the HCP-YA dataset is made up of monozygotic (MZ) and dizygotic (DZ) twin pairs (see \textit{Table \ref{table_numconns}}), so the next step was to utilize this and quantify the twin-fingerprint in the dataset which has not been done before. In this analysis, we adapt the identifiability framework by using, for each fMRI condition, one single-session functional connectome from each of the twins in a pair (MZ and DZ, separately) in lieu of test and retest of the same subject. We found the presence of a twin-fingerprint (both for MZ and DZ twins) in all fMRI conditions. It is noteworthy that the twin-fingerprint was much higher than expected by chance, at the same time being lower than individual fingerprint (based on test/retest Unrelated subjects; see \textit{Figure \ref{fig_continuum}}). Similar to the cohort of Unrelated subjects, the identifiability framework not only contributed in uncovering higher twin-fingerprints in all fMRI conditions, but also enabled us to utilize the higher granularity of the parcellations to achieve higher $I_{diff}$ scores. In particular, the $I_{diff}$ profiles are similar across the three cohorts, with the cohort of Unrelated subjects achieving the highest peaks, followed by MZ and DZ twins, respectively (see \textit{Figures \ref{fig_unrelated}, \ref{fig_MZ_wb}, \ref{fig_DZ_wb}}, and \textit{\ref{fig_continuum}}). This ordinal structure of the fingerprint across the three cohorts (namely Unrelated subjects, MZ twins,  and DZ twins) can be explained in terms of the genetic and environmental similarity between the pairs of connectomes across two sessions. In particular, for Unrelated subjects, the genetic information is 100\% equal and the environment is also highly shared across sessions as the scans belong to the same subject. On the other hand, the MZ twins, even though they share 100\% of their genetic information, they environment is shared to a much lower degree as the MZ twins are two separate individuals. Lastly, DZ twins share (on average) 50\% of their genetic information and the environment shared is similar to that of MZ twins \cite{koeppen2003twins}. Please note that, in \textit{Figure \ref{fig_continuum}}, we selected a subset of the Unrelated subjects and MZ twins cohorts in order to match number to DZ twins (69 pairs) in order to facilitate a meaningful comparison. \par
    
    The presence of a substantial twin-fingerprint (MZ and DZ) is a compelling argument for utilizing only a cohort of unrelated subjects when conducting studies that rely on brain fingerprinting and differences between individuals. This is because including twin pairs or siblings in such studies can confound the results, as evidenced by the findings of this paper. An alternative strategy could be to keep both twins from a pair either in the training or the validation dataset. If one of the twins is used for training and the other for validation, it might lead to a false increase in the prediction accuracy of the model under consideration \cite{seguin2020network}. \par
    
    In their seminal work, Finn et al., 2015 \cite{finn2015functional} observed that some of the functional networks of the brain contained a higher fingerprint than the whole-brain fingerprint at rest and between fMRI tasks (used as test/retest). In order to replicate and extend on this result, we quantified the amount of individual fingerprint for the 7 resting-state networks (or RSNs, as proposed by Yeo et al., 2011 \cite{yeo2011organization}) using the (single-session) resting-state fMRI condition across all available levels of granularity. Consistent with previous findings, \textit{Figure \ref{fig_idiff_yeo}} shows that some of the RSNs achieve a higher (e.g., somatomotor and frontoparietal) fingerprint than others (e.g., visual and limbic) using original FCs. Similar to the whole-brain scenarios, the identifiability framework uncovers the individual fingerprint in all the RSNs and enables us extract the higher fingerprint present in the higher granularity of the parcellations. Another effect of using the identifiability framework is that the amount of fingerprint present in each RSN becomes much more uniform, with all RSNs (except visual) reaching around $I_{diff}$ $\approx$ 35 for Schaefer900. \par
    
    Lastly, \textit{Figure \ref{fig_reduced_tseries}} shows the effect of scanning length on the identifiability of resting-state FCs. We have chosen to run this experiment only on resting-state data as it is the longest fMRI acquisition in the HCP-YA dataset. As can be seen from the profiles of original $I_{diff}$ scores for all the Schaefer parcellations, the difference between the identifiability of different parcellations is negligible up to 300 timepoints. The same is not true for the optimal $I_{diff}$, as the profiles start diverging from each other as early as 150 timepoints. The optimal $I_{diff}$ achieved is also higher than the original $I_{diff}$ at every scanning length, but the difference is more pronounced for higher granularity parcellations. Overall, we observe that a longer scanning duration leads to a higher fingerprint, but saturating at around $\sim$13 minutes for resting-state fMRI. In addition, identifiability framework not only allows us to uncover higher fingerprint for the same scanning duration, but also enables us to utilize the higher granularity parcellation to achieve higher fingerprint. \par
    
    Amongst the limitations of this work is the fact that a dataset such as HCP-YA inherently does not have a large cohort of twin subjects or different age groups. This has been a limitation in terms of not being able to ask specific research questions to study the fingerprint between twins or across the lifespan. Also, specific to HCP-YA, the task fMRI acquisition lengths are heterogeneous and not as long as resting state acquisition (see \textit{Table \ref{table_task_summary}}). The availability of longer task-based fMRI sequences would allow researchers to analyze the effect of scanning length and TR on the stability of the connectomes, similar to the analysis done on resting state fMRI data carried out in \textit{Sections \ref{sec_scanlen_effect}}. Lastly, we have not studied the effect of different processing pipelines on identifiability measures \cite{aquino2020identifying}\cite{vytvarova2017impact}. \par
    
    In the future, similar efforts could be pursued to also process the diffusion weighted imaging data included in the Human Connectome Project Young Adult dataset and make it available for public use of the corresponding subject-level structural connectomes. Researchers could also provide processed versions other state-of-the-art brain connectivity datasets such as Alzheimer's Disease Neuroimaging Initiative (ADNI) \cite{petersen2010alzheimer}, Adolescent Brain Cognitive Development (ABCD), HCP--Lifespan, HCP--Aging, among others together with their corresponding fingerprint analyses. \par
    
\section*{Competing Interest}
None of the authors have any financial competing interest related to this work.
    
\section*{Acknowledgement}

    Data were provided [in part] by the Human Connectome Project, WU-Minn Consortium (Principal Investigators: David Van Essen and Kamil Ugurbil; 1U54MH091657) funded by the 16 NIH Institutes and Centers that support the NIH Blueprint for Neuroscience Research; and by the McDonnell Center for Systems Neuroscience at Washington University. This work was supported in part by NIH R01 EB022574. \par
    
    We would like to thank the Lawrence Livermore National Laboratory Data Science Institute’s Open Data Initiative and Rushil Anirudh for coordinating the data release, along with the University of California, San Diego Library Digital Collections for hosting the data. Lawrence Livermore National Laboratory is operated by Lawrence Livermore National Security, LLC, for the U.S. Department of Energy, National Nuclear Security Administration under Contract DE-AC52-07NA27344. \par

    Authors acknowledge financial support from NIH R01EB022574 (JG and LS), NIH R01MH108467 (JG), Indiana Alcohol Research Center P60AA07611 (JG), Purdue Discovery Park Data Science Award "Finger- prints of the Human Brain: A Data Science Perspective ”(JG), and from the SNSF Ambizione project PZ00P2\_185716 "Fingeprinting the brain: network science to extract features of cognition, behavior and dysfunction" (EA).

\bibliographystyle{unsrt}
\bibliography{citations}

\begin{thebibliography}{10}

\bibitem{van2013wu}
David~C Van~Essen, Stephen~M Smith, Deanna~M Barch, Timothy~EJ Behrens, Essa
  Yacoub, Kamil Ugurbil, Wu-Minn~HCP Consortium, et~al.
\newblock The wu-minn human connectome project: an overview.
\newblock {\em Neuroimage}, 80:62--79, 2013.

\bibitem{bookheimer2019lifespan}
Susan~Y Bookheimer, David~H Salat, Melissa Terpstra, Beau~M Ances, Deanna~M
  Barch, Randy~L Buckner, Gregory~C Burgess, Sandra~W Curtiss, Mirella
  Diaz-Santos, Jennifer~Stine Elam, et~al.
\newblock The lifespan human connectome project in aging: an overview.
\newblock {\em NeuroImage}, 185:335--348, 2019.

\bibitem{somerville2018lifespan}
Leah~H Somerville, Susan~Y Bookheimer, Randy~L Buckner, Gregory~C Burgess,
  Sandra~W Curtiss, Mirella Dapretto, Jennifer~Stine Elam, Michael~S Gaffrey,
  Michael~P Harms, Cynthia Hodge, et~al.
\newblock The lifespan human connectome project in development: A large-scale
  study of brain connectivity development in 5--21 year olds.
\newblock {\em Neuroimage}, 183:456--468, 2018.

\bibitem{allen2014uk}
Naomi~E Allen, Cathie Sudlow, Tim Peakman, Rory Collins, et~al.
\newblock Uk biobank data: come and get it, 2014.

\bibitem{miller2016multimodal}
Karla~L Miller, Fidel Alfaro-Almagro, Neal~K Bangerter, David~L Thomas, Essa
  Yacoub, Junqian Xu, Andreas~J Bartsch, Saad Jbabdi, Stamatios~N Sotiropoulos,
  Jesper~LR Andersson, et~al.
\newblock Multimodal population brain imaging in the uk biobank prospective
  epidemiological study.
\newblock {\em Nature neuroscience}, 19(11):1523--1536, 2016.

\bibitem{petersen2010alzheimer}
Ronald~Carl Petersen, PS~Aisen, Laurel~A Beckett, MC~Donohue, AC~Gamst,
  Danielle~J Harvey, CR~Jack, WJ~Jagust, LM~Shaw, AW~Toga, et~al.
\newblock Alzheimer's disease neuroimaging initiative (adni): clinical
  characterization.
\newblock {\em Neurology}, 74(3):201--209, 2010.

\bibitem{glasser2013minimal}
Matthew~F Glasser, Stamatios~N Sotiropoulos, J~Anthony Wilson, Timothy~S
  Coalson, Bruce Fischl, Jesper~L Andersson, Junqian Xu, Saad Jbabdi, Matthew
  Webster, Jonathan~R Polimeni, et~al.
\newblock The minimal preprocessing pipelines for the human connectome project.
\newblock {\em Neuroimage}, 80:105--124, 2013.

\bibitem{makropoulos2018developing}
Antonios Makropoulos, Emma~C Robinson, Andreas Schuh, Robert Wright, Sean
  Fitzgibbon, Jelena Bozek, Serena~J Counsell, Johannes Steinweg, Katy
  Vecchiato, Jonathan Passerat-Palmbach, et~al.
\newblock The developing human connectome project: A minimal processing
  pipeline for neonatal cortical surface reconstruction.
\newblock {\em Neuroimage}, 173:88--112, 2018.

\bibitem{power2020critical}
Jonathan~D Power, Charles~J Lynch, Babatunde Adeyemo, and Steven~E Petersen.
\newblock A critical, event-related appraisal of denoising in resting-state
  fmri studies.
\newblock {\em Cerebral Cortex}, 2020.

\bibitem{parkes2018evaluation}
Linden Parkes, Ben Fulcher, Murat Y{\"u}cel, and Alex Fornito.
\newblock An evaluation of the efficacy, reliability, and sensitivity of motion
  correction strategies for resting-state functional mri.
\newblock {\em Neuroimage}, 171:415--436, 2018.

\bibitem{power2018ridding}
Jonathan~D Power, Mark Plitt, Stephen~J Gotts, Prantik Kundu, Valerie Voon,
  Peter~A Bandettini, and Alex Martin.
\newblock Ridding fmri data of motion-related influences: Removal of signals
  with distinct spatial and physical bases in multiecho data.
\newblock {\em Proceedings of the National Academy of Sciences},
  115(9):E2105--E2114, 2018.

\bibitem{power2014methods}
Jonathan~D Power, Anish Mitra, Timothy~O Laumann, Abraham~Z Snyder, Bradley~L
  Schlaggar, and Steven~E Petersen.
\newblock Methods to detect, characterize, and remove motion artifact in
  resting state fmri.
\newblock {\em Neuroimage}, 84:320--341, 2014.

\bibitem{power2015recent}
Jonathan~D Power, Bradley~L Schlaggar, and Steven~E Petersen.
\newblock Recent progress and outstanding issues in motion correction in
  resting state fmri.
\newblock {\em Neuroimage}, 105:536--551, 2015.

\bibitem{burgess2016evaluation}
Gregory~C Burgess, Sridhar Kandala, Dan Nolan, Timothy~O Laumann, Jonathan~D
  Power, Babatunde Adeyemo, Michael~P Harms, Steven~E Petersen, and Deanna~M
  Barch.
\newblock Evaluation of denoising strategies to address motion-correlated
  artifacts in resting-state functional magnetic resonance imaging data from
  the human connectome project.
\newblock {\em Brain connectivity}, 6(9):669--680, 2016.

\bibitem{fornito2016fundamentals}
Alex Fornito, Andrew Zalesky, and Edward Bullmore.
\newblock {\em Fundamentals of brain network analysis}.
\newblock Academic Press, 2016.

\bibitem{sporns2010networks}
Olaf Sporns.
\newblock {\em Networks of the Brain}.
\newblock MIT press, 2010.

\bibitem{sporns2012discovering}
Olaf Sporns.
\newblock {\em Discovering the human connectome}.
\newblock MIT press, 2012.

\bibitem{schaefer2018local}
Alexander Schaefer, Ru~Kong, Evan~M Gordon, Timothy~O Laumann, Xi-Nian Zuo,
  Avram~J Holmes, Simon~B Eickhoff, and BT~Thomas Yeo.
\newblock Local-global parcellation of the human cerebral cortex from intrinsic
  functional connectivity mri.
\newblock {\em Cerebral Cortex}, 28(9):3095--3114, 2018.

\bibitem{glasser2016multi}
Matthew~F Glasser, Timothy~S Coalson, Emma~C Robinson, Carl~D Hacker, John
  Harwell, Essa Yacoub, Kamil Ugurbil, Jesper Andersson, Christian~F Beckmann,
  Mark Jenkinson, et~al.
\newblock A multi-modal parcellation of human cerebral cortex.
\newblock {\em Nature}, 536(7615):171--178, 2016.

\bibitem{salehi2020there}
Mehraveh Salehi, Abigail~S Greene, Amin Karbasi, Xilin Shen, Dustin Scheinost,
  and R~Todd Constable.
\newblock There is no single functional atlas even for a single individual:
  Functional parcel definitions change with task.
\newblock {\em NeuroImage}, 208:116366, 2020.

\bibitem{finn2015functional}
Emily~S Finn, Xilin Shen, Dustin Scheinost, Monica~D Rosenberg, Jessica Huang,
  Marvin~M Chun, Xenophon Papademetris, and R~Todd Constable.
\newblock Functional connectome fingerprinting: identifying individuals using
  patterns of brain connectivity.
\newblock {\em Nature neuroscience}, 18(11):1664--1671, 2015.

\bibitem{abbas2020regularization}
Kausar Abbas, Mintao Liu, Manasij Venkatesh, Enrico Amico, Jaroslaw Harezlak,
  Alan~David Kaplan, Mario Ventresca, Luiz Pessoa, and Joaqu{\'\i}n Go{\~n}i.
\newblock Regularization of functional connectomes and its impact on geodesic
  distance and fingerprinting.
\newblock {\em arXiv preprint arXiv:2003.05393}, 2020.

\bibitem{yeo2011organization}
BT~Thomas Yeo, Fenna~M Krienen, Jorge Sepulcre, Mert~R Sabuncu, Danial
  Lashkari, Marisa Hollinshead, Joshua~L Roffman, Jordan~W Smoller, Lilla
  Z{\"o}llei, Jonathan~R Polimeni, et~al.
\newblock The organization of the human cerebral cortex estimated by intrinsic
  functional connectivity.
\newblock {\em Journal of neurophysiology}, 2011.

\bibitem{amico2018quest}
Enrico Amico and Joaqu{\'\i}n Go{\~n}i.
\newblock The quest for identifiability in human functional connectomes.
\newblock {\em Scientific reports}, 8(1):1--14, 2018.

\bibitem{abbas2020geff}
Kausar Abbas, Enrico Amico, Diana~Otero Svaldi, Uttara Tipnis, Duy~Anh
  Duong-Tran, Mintao Liu, Meenusree Rajapandian, Jaroslaw Harezlak, Beau~M
  Ances, and Joaqu{\'\i}n Go{\~n}i.
\newblock Geff: Graph embedding for functional fingerprinting.
\newblock {\em NeuroImage}, page 117181, 2020.

\bibitem{kaufmann2017delayed}
Tobias Kaufmann, Dag Aln{\ae}s, Nhat~Trung Doan, Christine~Lycke Brandt, Ole~A
  Andreassen, and Lars~T Westlye.
\newblock Delayed stabilization and individualization in connectome development
  are related to psychiatric disorders.
\newblock {\em Nature neuroscience}, 20(4):513--515, 2017.

\bibitem{miranda2014connectotyping}
Oscar Miranda-Dominguez, Brian~D Mills, Samuel~D Carpenter, Kathleen~A Grant,
  Christopher~D Kroenke, Joel~T Nigg, and Damien~A Fair.
\newblock Connectotyping: model based fingerprinting of the functional
  connectome.
\newblock {\em PloS one}, 9(11):e111048, 2014.

\bibitem{noble2017influences}
Stephanie Noble, Marisa~N Spann, Fuyuze Tokoglu, Xilin Shen, R~Todd Constable,
  and Dustin Scheinost.
\newblock Influences on the test--retest reliability of functional connectivity
  mri and its relationship with behavioral utility.
\newblock {\em Cerebral Cortex}, 27(11):5415--5429, 2017.

\bibitem{rajapandian2020uncovering}
Meenusree Rajapandian, Enrico Amico, Kausar Abbas, Mario Ventresca, and
  Joaqu{\'\i}n Go{\~n}i.
\newblock Uncovering differential identifiability in network properties of
  human brain functional connectomes.
\newblock {\em Network Neuroscience}, 4(3):698--713, 2020.

\bibitem{mars2018connectivity}
Rogier~B Mars, Richard~E Passingham, and Saad Jbabdi.
\newblock Connectivity fingerprints: from areal descriptions to abstract
  spaces.
\newblock {\em Trends in cognitive sciences}, 22(11):1026--1037, 2018.

\bibitem{hu2020disentangled}
Dan Hu, Fan Wang, Han Zhang, Zhengwang Wu, Li~Wang, Weili Lin, Gang Li,
  Dinggang Shen, UNC/UMN Baby Connectome~Project Consortium, et~al.
\newblock Disentangled intensive triplet autoencoder for infant functional
  connectome fingerprinting.
\newblock In {\em International Conference on Medical Image Computing and
  Computer-Assisted Intervention}, pages 72--82. Springer, 2020.

\bibitem{ngo2020connectomic}
Gia~H Ngo, Meenakshi Khosla, Keith Jamison, Amy Kuceyeski, and Mert~R Sabuncu.
\newblock From connectomic to task-evoked fingerprints: Individualized
  prediction of task contrasts from resting-state functional connectivity.
\newblock In {\em International Conference on Medical Image Computing and
  Computer-Assisted Intervention}, pages 62--71. Springer, 2020.

\bibitem{svaldi2019optimizing}
Diana~O Svaldi, Joaqu{\'\i}n Go{\~n}i, Kausar Abbas, Enrico Amico, David~G
  Clark, Charanya Muralidharan, Mario Dzemidzic, John~D West, Shannon~L
  Risacher, Andrew~J Saykin, et~al.
\newblock Optimizing differential identifiability improves connectome
  predictive modeling of cognitive deficits in alzheimer$\backslash$'s disease.
\newblock {\em arXiv preprint arXiv:1908.06197}, 2019.

\bibitem{sripada2020boost}
Chandra Sripada, Aman Taxali, Mike Angstadt, and Saige Rutherford.
\newblock Boost in test-retest reliability in resting state fmri with
  predictive modeling.
\newblock {\em BioRxiv}, page 796714, 2020.

\bibitem{ge2017heritability}
Tian Ge, Avram~J Holmes, Randy~L Buckner, Jordan~W Smoller, and Mert~R Sabuncu.
\newblock Heritability analysis with repeat measurements and its application to
  resting-state functional connectivity.
\newblock {\em Proceedings of the National Academy of Sciences},
  114(21):5521--5526, 2017.

\bibitem{de2008electroencephalographic}
Luigi De~Gennaro, Cristina Marzano, Fabiana Fratello, Fabio Moroni,
  Maria~Concetta Pellicciari, Fabio Ferlazzo, Stefania Costa, Alessandro
  Couyoumdjian, Giuseppe Curcio, Emilia Sforza, et~al.
\newblock The electroencephalographic fingerprint of sleep is genetically
  determined: a twin study.
\newblock {\em Annals of neurology}, 64(4):455--460, 2008.

\bibitem{kumar2018multi}
Kuldeep Kumar, Matthew Toews, Laurent Chauvin, Olivier Colliot, and Christian
  Desrosiers.
\newblock Multi-modal brain fingerprinting: a manifold approximation based
  framework.
\newblock {\em NeuroImage}, 183:212--226, 2018.

\bibitem{gritsenko2020twin}
Andrey Gritsenko, Martin Lindquist, and Moo~K Chung.
\newblock Twin classification in resting-state brain connectivity.
\newblock In {\em 2020 IEEE 17th International Symposium on Biomedical Imaging
  (ISBI)}, pages 1391--1394. IEEE, 2020.

\bibitem{colclough2017heritability}
Giles~L Colclough, Stephen~M Smith, Thomas~E Nichols, Anderson~M Winkler,
  Stamatios~N Sotiropoulos, Matthew~F Glasser, David~C Van~Essen, and Mark~W
  Woolrich.
\newblock The heritability of multi-modal connectivity in human brain activity.
\newblock {\em Elife}, 6:e20178, 2017.

\bibitem{demeter2020functional}
Damion~V Demeter, Laura~E Engelhardt, Remington Mallett, Evan~M Gordon, Tehila
  Nugiel, K~Paige Harden, Elliot~M Tucker-Drob, Jarrod~A Lewis-Peacock, and
  Jessica~A Church.
\newblock Functional connectivity fingerprints at rest are similar across
  youths and adults and vary with genetic similarity.
\newblock {\em Iscience}, 23(1):100801, 2020.

\bibitem{rubinov2010complex}
Mikail Rubinov and Olaf Sporns.
\newblock Complex network measures of brain connectivity: uses and
  interpretations.
\newblock {\em Neuroimage}, 52(3):1059--1069, 2010.

\bibitem{murphy2017towards}
Kevin Murphy and Michael~D Fox.
\newblock Towards a consensus regarding global signal regression for resting
  state functional connectivity mri.
\newblock {\em Neuroimage}, 154:169--173, 2017.

\bibitem{liu2017global}
Thomas~T Liu, Alican Nalci, and Maryam Falahpour.
\newblock The global signal in fmri: Nuisance or information?
\newblock {\em NeuroImage}, 150:213--229, 2017.

\bibitem{hayasaka2013functional}
Satoru Hayasaka.
\newblock Functional connectivity networks with and without global signal
  correction.
\newblock {\em Frontiers in human neuroscience}, 7:880, 2013.

\bibitem{xu2018impact}
Huaze Xu, Jianpo Su, Jian Qin, Ming Li, Ling-Li Zeng, Dewen Hu, and Hui Shen.
\newblock Impact of global signal regression on characterizing dynamic
  functional connectivity and brain states.
\newblock {\em Neuroimage}, 173:127--145, 2018.

\bibitem{gotts2013perils}
Stephen~J Gotts, Ziad~S Saad, Hang~Joon Jo, Gregory~L Wallace, Robert~W Cox,
  and Alex Martin.
\newblock The perils of global signal regression for group comparisons: a case
  study of autism spectrum disorders.
\newblock {\em Frontiers in human neuroscience}, 7:356, 2013.

\bibitem{saad2012trouble}
Ziad~S Saad, Stephen~J Gotts, Kevin Murphy, Gang Chen, Hang~Joon Jo, Alex
  Martin, and Robert~W Cox.
\newblock Trouble at rest: how correlation patterns and group differences
  become distorted after global signal regression.
\newblock {\em Brain connectivity}, 2(1):25--32, 2012.

\bibitem{bari2019uncovering}
Sumra Bari, Enrico Amico, Nicole Vike, Thomas~M Talavage, and Joaqu{\'\i}n
  Go{\~n}i.
\newblock Uncovering multi-site identifiability based on resting-state
  functional connectomes.
\newblock {\em NeuroImage}, 202:115967, 2019.

\bibitem{van2012human}
David~C Van~Essen, Kamil Ugurbil, E~Auerbach, D~Barch, TEJ Behrens, R~Bucholz,
  Acer Chang, Liyong Chen, Maurizio Corbetta, Sandra~W Curtiss, et~al.
\newblock The human connectome project: a data acquisition perspective.
\newblock {\em Neuroimage}, 62(4):2222--2231, 2012.

\bibitem{barch2013function}
Deanna~M Barch, Gregory~C Burgess, Michael~P Harms, Steven~E Petersen,
  Bradley~L Schlaggar, Maurizio Corbetta, Matthew~F Glasser, Sandra Curtiss,
  Sachin Dixit, Cindy Feldt, et~al.
\newblock Function in the human connectome: task-fmri and individual
  differences in behavior.
\newblock {\em Neuroimage}, 80:169--189, 2013.

\bibitem{hariri2006preference}
Ahmad~R Hariri, Sarah~M Brown, Douglas~E Williamson, Janine~D Flory, Harriet
  De~Wit, and Stephen~B Manuck.
\newblock Preference for immediate over delayed rewards is associated with
  magnitude of ventral striatal activity.
\newblock {\em Journal of Neuroscience}, 26(51):13213--13217, 2006.

\bibitem{delgado2000tracking}
Mauricio~R Delgado, Leigh~E Nystrom, Catherine Fissell, DC~Noll, and Julie~A
  Fiez.
\newblock Tracking the hemodynamic responses to reward and punishment in the
  striatum.
\newblock {\em Journal of neurophysiology}, 84(6):3072--3077, 2000.

\bibitem{binder2011mapping}
Jeffrey~R Binder, William~L Gross, Jane~B Allendorfer, Leonardo Bonilha,
  Jessica Chapin, Jonathan~C Edwards, Thomas~J Grabowski, John~T Langfitt,
  David~W Loring, Mark~J Lowe, et~al.
\newblock Mapping anterior temporal lobe language areas with fmri: a
  multicenter normative study.
\newblock {\em Neuroimage}, 54(2):1465--1475, 2011.

\bibitem{smith2007localizing}
Rachelle Smith, Kamyar Keramatian, and Kalina Christoff.
\newblock Localizing the rostrolateral prefrontal cortex at the individual
  level.
\newblock {\em Neuroimage}, 36(4):1387--1396, 2007.

\bibitem{castelli2000movement}
Fulvia Castelli, Francesca Happ{\'e}, Uta Frith, and Chris Frith.
\newblock Movement and mind: a functional imaging study of perception and
  interpretation of complex intentional movement patterns.
\newblock {\em Neuroimage}, 12(3):314--325, 2000.

\bibitem{castelli2002autism}
Fulvia Castelli, Chris Frith, Francesca Happ{\'e}, and Uta Frith.
\newblock Autism, asperger syndrome and brain mechanisms for the attribution of
  mental states to animated shapes.
\newblock {\em Brain}, 125(8):1839--1849, 2002.

\bibitem{wheatley2007understanding}
Thalia Wheatley, Shawn~C Milleville, and Alex Martin.
\newblock Understanding animate agents: distinct roles for the social network
  and mirror system.
\newblock {\em Psychological science}, 18(6):469--474, 2007.

\bibitem{white2011developing}
Sarah~J White, Devorah Coniston, Rosannagh Rogers, and Uta Frith.
\newblock Developing the frith-happ{\'e} animations: A quick and objective test
  of theory of mind for adults with autism.
\newblock {\em Autism Research}, 4(2):149--154, 2011.

\bibitem{drobyshevsky2006rapid}
Alexander Drobyshevsky, Stephen~B Baumann, and Walter Schneider.
\newblock A rapid fmri task battery for mapping of visual, motor, cognitive,
  and emotional function.
\newblock {\em Neuroimage}, 31(2):732--744, 2006.

\bibitem{caceres2009measuring}
Alejandro Caceres, Deanna~L Hall, Fernando~O Zelaya, Steven~CR Williams, and
  Mitul~A Mehta.
\newblock Measuring fmri reliability with the intra-class correlation
  coefficient.
\newblock {\em Neuroimage}, 45(3):758--768, 2009.

\bibitem{downing2001cortical}
Paul~E Downing, Yuhong Jiang, Miles Shuman, and Nancy Kanwisher.
\newblock A cortical area selective for visual processing of the human body.
\newblock {\em Science}, 293(5539):2470--2473, 2001.

\bibitem{peelen2005within}
Marius~V Peelen and Paul~E Downing.
\newblock Within-subject reproducibility of category-specific visual activation
  with functional mri.
\newblock {\em Human brain mapping}, 25(4):402--408, 2005.

\bibitem{taylor2007functional}
John~C Taylor, Alison~J Wiggett, and Paul~E Downing.
\newblock Functional mri analysis of body and body part representations in the
  extrastriate and fusiform body areas.
\newblock {\em Journal of neurophysiology}, 98(3):1626--1633, 2007.

\bibitem{fox2009defining}
Christopher~J Fox, Giuseppe Iaria, and Jason~JS Barton.
\newblock Defining the face processing network: optimization of the functional
  localizer in fmri.
\newblock {\em Human brain mapping}, 30(5):1637--1651, 2009.

\bibitem{kung2007region}
Chun-Chia Kung, Jessie~J Peissig, and Michael~J Tarr.
\newblock Is region-of-interest overlap comparison a reliable measure of
  category specificity?
\newblock {\em Journal of Cognitive Neuroscience}, 19(12):2019--2034, 2007.

\bibitem{smith2013resting}
Stephen~M Smith, Christian~F Beckmann, Jesper Andersson, Edward~J Auerbach,
  Janine Bijsterbosch, Gwena{\"e}lle Douaud, Eugene Duff, David~A Feinberg,
  Ludovica Griffanti, Michael~P Harms, et~al.
\newblock Resting-state fmri in the human connectome project.
\newblock {\em Neuroimage}, 80:144--168, 2013.

\bibitem{van2012parcellations}
David~C Van~Essen, Matthew~F Glasser, Donna~L Dierker, John Harwell, and
  Timothy Coalson.
\newblock Parcellations and hemispheric asymmetries of human cerebral cortex
  analyzed on surface-based atlases.
\newblock {\em Cerebral cortex}, 22(10):2241--2262, 2012.

\bibitem{salimi2014automatic}
Gholamreza Salimi-Khorshidi, Gwena{\"e}lle Douaud, Christian~F Beckmann,
  Matthew~F Glasser, Ludovica Griffanti, and Stephen~M Smith.
\newblock Automatic denoising of functional mri data: combining independent
  component analysis and hierarchical fusion of classifiers.
\newblock {\em Neuroimage}, 90:449--468, 2014.

\bibitem{griffanti2014ica}
Ludovica Griffanti, Gholamreza Salimi-Khorshidi, Christian~F Beckmann, Edward~J
  Auerbach, Gwena{\"e}lle Douaud, Claire~E Sexton, Enik{\H{o}} Zsoldos, Klaus~P
  Ebmeier, Nicola Filippini, Clare~E Mackay, et~al.
\newblock Ica-based artefact removal and accelerated fmri acquisition for
  improved resting state network imaging.
\newblock {\em Neuroimage}, 95:232--247, 2014.

\bibitem{aquino2020identifying}
Kevin~M Aquino, Ben~D Fulcher, Linden Parkes, Kristina Sabaroedin, and Alex
  Fornito.
\newblock Identifying and removing widespread signal deflections from fmri
  data: Rethinking the global signal regression problem.
\newblock {\em NeuroImage}, 212:116614, 2020.

\bibitem{prescott1995twin}
Carol~A Prescott and Kenneth~S Kendler.
\newblock Twin study design.
\newblock {\em Alcohol Health and Research World}, 19(3):200, 1995.

\bibitem{blokland2013twin}
Gabri{\"e}lla~AM Blokland, Miriam~A Mosing, Karin~H Verweij, and Sarah~E
  Medland.
\newblock Twin studies and behavior genetics.
\newblock {\em The Oxford handbook of quantitative methods in psychology},
  2:198--218, 2013.

\bibitem{deng2013nonlinear}
Xiaogang Deng and Xuemin Tian.
\newblock Nonlinear process fault pattern recognition using statistics kernel
  pca similarity factor.
\newblock {\em Neurocomputing}, 121:298--308, 2013.

\bibitem{hsieh2009novel}
Ping-Cheng Hsieh and Pi-Cheng Tung.
\newblock A novel hybrid approach based on sub-pattern technique and whitened
  pca for face recognition.
\newblock {\em Pattern Recognition}, 42(5):978--984, 2009.

\bibitem{manjon2013diffusion}
Jos{\'e}~V Manj{\'o}n, Pierrick Coup{\'e}, Luis Concha, Antonio Buades, D~Louis
  Collins, and Montserrat Robles.
\newblock Diffusion weighted image denoising using overcomplete local pca.
\newblock {\em PloS one}, 8(9):e73021, 2013.

\bibitem{de2007denoising}
Alain De~Cheveign{\'e} and Jonathan~Z Simon.
\newblock Denoising based on time-shift pca.
\newblock {\em Journal of neuroscience methods}, 165(2):297--305, 2007.

\bibitem{fisher1915frequency}
Ronald~A Fisher.
\newblock Frequency distribution of the values of the correlation coefficient
  in samples from an indefinitely large population.
\newblock {\em Biometrika}, 10(4):507--521, 1915.

\bibitem{negishi2011functional}
Michiro Negishi, Roberto Martuzzi, Edward~J Novotny, Dennis~D Spencer, and
  R~Todd Constable.
\newblock Functional mri connectivity as a predictor of the surgical outcome of
  epilepsy.
\newblock {\em Epilepsia}, 52(9):1733--1740, 2011.

\bibitem{hampson2010functional}
Michelle Hampson, Naomi Driesen, Jennifer~K Roth, John~C Gore, and R~Todd
  Constable.
\newblock Functional connectivity between task-positive and task-negative brain
  areas and its relation to working memory performance.
\newblock {\em Magnetic resonance imaging}, 28(8):1051--1057, 2010.

\bibitem{tomasi2012resting}
Dardo Tomasi and Nora~D Volkow.
\newblock Resting functional connectivity of language networks:
  characterization and reproducibility.
\newblock {\em Molecular psychiatry}, 17(8):841--854, 2012.

\bibitem{fox2013identification}
Michael~D Fox, Hesheng Liu, and Alvaro Pascual-Leone.
\newblock Identification of reproducible individualized targets for treatment
  of depression with tms based on intrinsic connectivity.
\newblock {\em Neuroimage}, 66:151--160, 2013.

\bibitem{mark2017using}
Katharine~M Mark, Alison Pike, Rachel~M Latham, and Bonamy~R Oliver.
\newblock Using twins to better understand sibling relationships.
\newblock {\em Behavior genetics}, 47(2):202--214, 2017.

\bibitem{pallares2018extracting}
Vicente Pallar{\'e}s, Andrea Insabato, Ana Sanju{\'a}n, Simone K{\"u}hn, Dante
  Mantini, Gustavo Deco, and Matthieu Gilson.
\newblock Extracting orthogonal subject-and condition-specific signatures from
  fmri data using whole-brain effective connectivity.
\newblock {\em Neuroimage}, 178:238--254, 2018.

\bibitem{liu2018chronnectome}
Jin Liu, Xuhong Liao, Mingrui Xia, and Yong He.
\newblock Chronnectome fingerprinting: Identifying individuals and predicting
  higher cognitive functions using dynamic brain connectivity patterns.
\newblock {\em Human brain mapping}, 39(2):902--915, 2018.

\bibitem{byrge2019high}
Lisa Byrge and Daniel~P Kennedy.
\newblock High-accuracy individual identification using a “thin slice” of
  the functional connectome.
\newblock {\em Network Neuroscience}, 3(2):363--383, 2019.

\bibitem{mueller2013individual}
Sophia Mueller, Danhong Wang, Michael~D Fox, BT~Thomas Yeo, Jorge Sepulcre,
  Mert~R Sabuncu, Rebecca Shafee, Jie Lu, and Hesheng Liu.
\newblock Individual variability in functional connectivity architecture of the
  human brain.
\newblock {\em Neuron}, 77(3):586--595, 2013.

\bibitem{faskowitz2020edge}
Joshua Faskowitz, Farnaz~Zamani Esfahlani, Youngheun Jo, Olaf Sporns, and
  Richard~F Betzel.
\newblock Edge-centric functional network representations of human cerebral
  cortex reveal overlapping system-level architecture.
\newblock Technical report, Nature Publishing Group, 2020.

\bibitem{venkatesh2020comparing}
Manasij Venkatesh, Joseph Jaja, and Luiz Pessoa.
\newblock Comparing functional connectivity matrices: A geometry-aware approach
  applied to participant identification.
\newblock {\em NeuroImage}, 207:116398, 2020.

\bibitem{satterthwaite2018personalized}
Theodore~D Satterthwaite, Cedric~H Xia, and Danielle~S Bassett.
\newblock Personalized neuroscience: Common and individual-specific features in
  functional brain networks.
\newblock {\em Neuron}, 98(2):243--245, 2018.

\bibitem{gratton2018functional}
Caterina Gratton, Timothy~O Laumann, Ashley~N Nielsen, Deanna~J Greene, Evan~M
  Gordon, Adrian~W Gilmore, Steven~M Nelson, Rebecca~S Coalson, Abraham~Z
  Snyder, Bradley~L Schlaggar, et~al.
\newblock Functional brain networks are dominated by stable group and
  individual factors, not cognitive or daily variation.
\newblock {\em Neuron}, 98(2):439--452, 2018.

\bibitem{koeppen2003twins}
Gesina Koeppen-Schomerus, Frank~M Spinath, and Robert Plomin.
\newblock Twins and non-twin siblings: Different estimates of shared
  environmental influence in early childhood.
\newblock {\em Twin Research and Human Genetics}, 6(2):97--105, 2003.

\bibitem{seguin2020network}
Caio Seguin, Ye~Tian, and Andrew Zalesky.
\newblock Network communication models improve the behavioral and functional
  predictive utility of the human structural connectome.
\newblock {\em bioRxiv}, 2020.

\bibitem{vytvarova2017impact}
Eva V{\`y}tvarov{\'a}, Jan Fousek, Marek Barto{\v{n}}, Radek Mare{\v{c}}ek,
  Martin Gajdo{\v{s}}, Martin Lamo{\v{s}}, Marie Nov{\'a}kov{\'a},
  Tom{\'a}{\v{s}} Slav{\'\i}{\v{c}}ek, Igor Peterlik, and Michal Mikl.
\newblock The impact of diverse preprocessing pipelines on brain functional
  connectivity.
\newblock In {\em 2017 25th European Signal Processing Conference (EUSIPCO)},
  pages 2644--2648. IEEE, 2017.

\end{thebibliography}

\end{document}